\documentclass[11pt,a4paper]{article}
\usepackage{jcappub}
\usepackage{comment}
\usepackage[english]{babel} 
\usepackage{graphicx} 
\usepackage[usenames,dvipsnames,svgnames,table]{xcolor}
\usepackage{commath}
\usepackage{bm}
\usepackage{amsfonts}
\usepackage{amsmath}
\usepackage{amssymb}
\usepackage{natbib}
\usepackage{appendix}
\usepackage{soul}
\usepackage[colorlinks = true,
            linkcolor = blue,
            urlcolor  = blue,
            citecolor = blue,
            anchorcolor = blue]{hyperref}

\usepackage[scale=0.8]{geometry}

\def\omk{{\Omega_{K,0}}}
\def\omo{{\Omega_{{\rm m},0}}}
\def\omc{{\Omega_{{\rm c},0}}}
\def\omb{{\Omega_{{\rm b},0}}}
\def\fsig{{f\sigma_{8}}}
\def\sig{{\sigma_{8}}}
\def\sig80{{\sigma_{8,0}}}

\newcommand{\fcal}{ {\mathcal F} }

\newcommand{\scal}{ {\mathcal S} }
\newcommand{\rpar}{r_{\parallel}}
\newcommand{\rper}{r_{\perp}}
\newcommand{\apar}{\alpha_{\parallel}}
\newcommand{\aper}{\alpha_{\perp}}

\begin{document}

\title{Constraining spatial curvature with large-scale structure}

\author[a]{Julien Bel,}
\author[b]{Julien Larena,}
\author[c,d,e]{Roy Maartens,}
\author[a]{Christian Marinoni,}
\author[c]{Louis Perenon}

\affiliation[a]{Centre de Physique Th\'eorique, Aix Marseille Universit\'e, Universit\'e de Toulon, CNRS, Marseille, France}
\affiliation[b]{Laboratoire Univers et Particules de Montpellier, Universit\'e de Montpellier, CNRS, Montpellier, France}
\affiliation[c]{Department of Physics \& Astronomy, University of the Western Cape, Cape Town 7535, South Africa}
\affiliation[d]{Institute of Cosmology \& Gravitation, University of Portsmouth, Portsmouth PO1 3FX, UK}
\affiliation[e]{National Institute for Theoretical \& Computational Sciences (NITheCS), Cape Town 7535, South Africa}

\emailAdd{julien.bel@cpt.univ-mrs.fr, julien.larena@umontpellier.fr, roy.maartens@gmail.com, christian.marinoni@cpt.univ-mrs.fr, perenon.louis@yahoo.fr}

\keywords{Cosmology, large-scale structure, spatial curvature}

\abstract{We analyse the clustering of matter on large scales in an extension of the concordance model that allows for spatial curvature. We develop a consistent approach to curvature and wide-angle effects on the galaxy 2-point correlation function in redshift space. In particular we derive the Alcock-Paczynski distortion of $f\sigma_{8}$, which differs significantly from empirical models in the literature. A key innovation is the use of the `Clustering Ratio', which probes clustering in a different way to redshift-space distortions, so that their combination delivers more powerful cosmological constraints. We use this combination to constrain cosmological parameters, without CMB information. In a curved Universe, we find that $\Omega_{{\rm m}, 0}=0.26\pm 0.04$ (68\% CL). When the clustering probes are combined with low-redshift background probes -- BAO and SNIa -- we obtain a CMB-independent constraint on curvature: $\Omega_{K,0} = 0.0041\,_{-0.0504}^{+0.0500}$. We find no Bayesian evidence that the flat concordance model can be rejected. In addition we show that the sound horizon at decoupling is $r_{\rm d} = 144.57 \pm 2.34 \; {\rm Mpc}$, in agreement with its measurement from CMB anisotropies. As a consequence, the late-time Universe is compatible with flat $\Lambda$CDM and a standard sound horizon, leading to a small value of $H_{0}$, {\em without} assuming any CMB information. Clustering Ratio measurements produce the only low-redshift clustering data set that is not in disagreement with the CMB, and combining the two data sets we obtain $\Omega_{K,0}= -0.023 \pm 0.010$.}

\maketitle

\section{Introduction}

Quantifying the abundance of the different forms of energy in the Universe as well as characterising their perturbations, are fundamental objectives of current and planned cosmological observations of the cosmic microwave background (CMB) and the large-scale structure. 
The consensus scenario that is currently supported by most observational data is a cosmological model containing, in addition to photons and neutrinos, standard (baryonic) matter, cold dark matter (CDM) and dark energy in the form of a cosmological constant $\Lambda$: the `concordance' spatially flat $\Lambda$CDM model. Adding interest to this landscape, already rich in implications for fundamental physics, are recent claims that spatial curvature of the Universe must be accounted for in the cosmic energy budget. 

Indeed, there is statistical evidence from the Planck experiment of a lensing contribution to the CMB power spectra that is anomalously large compared to what expected in the standard flat $\Lambda$CDM scenario -- and to what is directly measured using the Planck lensing-generated 4-point correlation function \cite{Planck:2018vyg}. As was highlighted in \cite{Planck:2018vyg}, positive curvature ($K>0,\, \Omega_K<0$), i.e. closed spatial sections, naturally explains away this tension. Indeed, the Planck data (TT, TE, EE+lowE power spectra) best fit a curved $\Lambda$CDM model (hereafter $K \Lambda$CDM) and suggest in particular that $-0.095 < \omk < -0.007$ with a probability of $99\%$ \cite{Planck:2018vyg}. Translated into Bayesian language, these results indicate a preference for spatially closed universes, with Bayesian betting probabilities of more than 50:1 against a flat universe \cite{Handley:2019tkm}, a conclusion confirmed also by the analysis of \cite{DiValentino:2019qzk}. However, in the absence of a well-motivated choice for the prior probability distribution of $\omk$, there is a danger of over-interpreting the posterior probability of the parameter \cite{Efstathiou:2020wem}. (Note that most analyses use a uniform, uninformative, distribution.)

At first glance, this result seems to suggest the possibility of new physics beyond the standard model, or even a crisis within the model itself. But it turns out that this scenario exacerbates the discrepancies between predictions and estimates for most local cosmological observables, such as the Hubble constant $H_0$ or the matter density parameter $\omo$ \cite{DiValentino:2019qzk}. Moreover, when Planck 2018 measurements are combined with external, low-redshift, datasets from supernova (SNIa) distances \cite{DiValentino:2019qzk}, cosmic chronometers \cite{Vagnozzi:2020dfn, Dhawan:2021mel}, the baryonic acoustic oscillation (BAO) scale \cite{Planck:2018vyg}, or clustering \cite{Chudaykin:2020ghx}, then spatial flatness is generally recovered. This is also supported by CMB data from other experiments, like the Atacama Cosmology Telescope (ACT) \cite{ACT:2020gnv}. The absence of supporting evidence from non-Planck observations is a further reason for caution about claims of a possible nonzero $\Omega_K$. A potential theoretical argument against a spatially curved model is that Inflation generically predicts a late-period cosmos that is extremely close to spatially flat \cite{Guth:1980zm}. However, $\Omega_K\to 0$ is not equivalent to $K=0$, and viable Inflation models can be constructed with $K\neq 0$ \cite{Hergt:2022fxk}) (see also \cite{Ellis:2020pis}).

Curved-space cosmologies have received little attention in the literature also because of the increased theoretical complexity involved in their analysis. Although it is straightforward to deal with the additional degrees of freedom when analysing the expansion kinematics of the cosmological background, quantifying its imprints on perturbations is highly nontrivial. {Pioneering work was done by \cite{Matsubara:1999du}, including a subtle Fourier analysis (for later work on curved perturbations, see e.g. \cite{DiDio:2016ykq})}. Regarding the primordial power spectrum, there is not a well-developed and accepted model for the origin of fluctuations in a spatially curved Universe. In particular, it is not obvious how to generalise the notion of scale-invariant fluctuations to scales where curvature becomes effective \cite{Efstathiou:2003hk,Handley:2019anl,Thavanesan:2020lov}. 

The next generation of spectroscopic surveys, such as those with DESI \cite{DESI:2016fyo}, Euclid \cite{Euclid:2019clj}, WFIRST \cite{Green:2012mj} and SKAO \cite{Bacon:2018dui}, will probe the Universe on ultra-large scales. With this exciting observational prospect, it becomes critical to understand, from a theoretical perspective, how the clustering of matter is affected by nonzero spatial curvature. On the ultra-large scales included in next-generation surveys, it is also necessary to incorporate relativistic light-cone effects on perturbations (see e.g. the early works \cite{Yoo:2010ni,Bonvin:2011bg,Challinor:2011bk,Jeong:2011as,Bertacca:2012tp} and see e.g. \cite{Viljoen:2021ocx} and references therein for more recent work). 

For data from current spectroscopic surveys, which do not include ultra-large scales, it is still important to develop models that go beyond the traditional zero-curvature and plane-parallel approximations. In this paper, we assess the impact of curvature and wide-angle corrections on large-scale structure correlations. Our goal is to contribute to building a consistent formalism that includes curvature and wide-angle effects and that removes implicit flatness assumptions from analysis algorithms.

On the observational side, multiple probes have been used to constrain $\omk$, either using unperturbed background observables, such as BAO and SNIa geometric probes, or using perturbed observables, such as the CMB lensing power spectrum and the galaxy power spectrum. There is evidence that, when these data are combined with CMB observations, the $68\%$ CL contours move back to the `natural place' which includes the parameters of the flat $\Lambda$CDM model. However, a key issue here is the correct statistical treatment of heterogeneous data. For example both \cite{Handley:2019tkm} and \cite{DiValentino:2019qzk} argue that within the assumption of non-flat spaces, such a combination of data sets should be viewed with caution, due to the mutual disagreement between the samples, i.e. the fact that they do not plausibly represent different accidental realisations of the same underlying Universe. 

These concerns are not just about combining CMB measurements with background probes such as BAO or SNIa, as already highlighted in \cite{DiValentino:2019qzk} -- but also with the measurement of the full-shape power spectrum \cite{Vagnozzi:2020rcz}. In this context, our paper aims to explore whether alternative clustering observables, namely redshift-space distortions (RSD) \cite{Howlett:2017asq, Huterer:2016uyq, Turnbull:2011ty, Hudson:2012gt, Davis:2010sw,Shi:2017qpr, Howlett:2014opa,Song:2008qt,Blake:2013nif, Samushia:2011cs, Wang:2017wia, Blake:2012pj, delaTorre:2016rxm, Pezzotta:2016gbo, Okumura:2015lvp, Zhao:2018gvb} (ranging over $0<z\lesssim2$) and the Clustering Ratio \cite{Zennaro:2017qnp} (ranging over $0.15<z< 0.67$) could help to resolve this impasse and put statistical inference on a more secure footing. A parallel goal is to investigate whether combining clustering and background probes of the low-redshift Universe can reduce uncertainties and provide curvature information without the need for CMB data. 

The paper is organised as follows. In \autoref{sec2} we revisit the theory of clustering in curved-space models. To this end, we simplify and unify the linear perturbation formalism originally developed by \cite{Matsubara:1999du}. We show the characteristic imprints of curvature in the low-redshift matter power spectrum, in multipoles of the 2-point correlation function and in the Clustering Ratio. This analysis includes wide-angle effects. In \autoref{sec:data} we discuss the methods and data sets we adopt. Special attention is given to analysing the transformation rule -- the Alcock-Paczynski (AP) correction \cite{Alcock:1979mp} -- which transforms the (model-dependent) RSD parameter $f \sigma_8$ predicted in a given cosmology into the value in the fiducial cosmology used to estimate it from the data. In particular, we show that the theoretically motivated prescription we derive deviates significantly from the standard phenomenological form typically adopted in the literature. Our results are presented and discussed in \autoref{sec4}. In particular, in \autoref{sec4.1} we show the constraints on the curved $\Lambda$CDM model ($K\Lambda$CDM) from low-redshift clustering probes. We further pursue this goal in \autoref{sec4.2}, by combining clustering measurements with geometric probes such as BAO and SNIa, and in \autoref{sec4.3}, by combining clustering with the CMB measurements. Special attention is given to present results obtained by combining samples that are consistent according to specific Bayesian criteria. Since our analysis does not provide any significant evidence to reject the standard concordance model, in \autoref{sec5} we present the constraints on the parameters of the flat $\Lambda$CDM model. Concluding remarks are provided in \autoref{sec6}. Technical details concerning the calculation of the AP distortion effects on both the RSD parameter $f\sigma_8$ and the Clustering Ratio are given in \autoref{apeffect} and \autoref{apcr}.

\section{Clustering observables in spatially curved models}\label{sec2}

The general homogeneous and isotropic FLRW metric is
\begin{equation}
\dif s^2= c^2\dif t^2 - a^2(t)\,\gamma_{ij}\dif x^i \dif x^j
= c^2\dif t^2 - a^2(t)\left [ \dif\chi^2 + S_K^2(\chi)\left ( \dif \theta^2 + \sin^2\theta\dif \phi^2 \right ) \right].
\label{metric}
\end{equation}
Here the comoving spatial coordinates $\chi,\theta,\phi$ are dimensionless and the scale factor $a$ has dimension length. The comoving line-of-sight distance to an event at time $t$ is defined as
\begin{eqnarray}
r=a_0\,\chi= ca_0 \int_{t_0}^{t}\frac{\dif t'}{a( t')}.
\label{distcom}
\end{eqnarray}
The curvature parameter $K$ has the values $-1,0,1$, for hyperbolic (open), flat (open), spherical (closed) spatial geometries, and
\begin{equation}\label{sk}
S_K(\chi) = \left \{ 
 \begin{array}{lcl}
  \sin\chi & \quad{\rm if} & K=1 \\
  \chi & \quad{\rm if} & K=0 \\
  \sinh\chi & \quad{\rm if} & K=-1
 \end{array}
 \right . 
\end{equation}
The angular distance $D_A$ is then 
\begin{eqnarray}
D_A = a\, S_K (\chi).
\end{eqnarray}
We also define $C_K(\chi)=\dif S_K(\chi)/\dif\chi$, so that $C_K^2(\chi)+K S_K^2(\chi)=1$. The dimensionless energy density associated with curvature is
\begin{equation}\label{omk}
\Omega_K = -{K c^2\over a^2 H^2}. 
\end{equation}
In a flat universe, by convention we use $a_0 = {c}/{H_0}$, while for curved models $a_0$ follows from \eqref{omk}:
\begin{eqnarray}\label{a0}
a_0= {c\over H_0}\,\left\{
\begin{array}{lc}
 1  & \quad K=0 \\
 \big| \omk\big|^{-1/2}  & \quad K\neq 0
\end{array}
\right.
\end{eqnarray}

\subsection{Real-space power spectrum}\label{sec2.1}

The global curvature of the universe changes the Laplacian operator and this affects the Fourier transform. The Fourier basis functions are plane waves in the flat case, but in general they are solutions of the Helmholtz equation
\begin{equation}
\tilde{\nabla}^2 \mathcal{Q}=\frac{1}{\sqrt{\gamma}}\,\partial_i \left ( \sqrt{\gamma}\,{\gamma^{ij}}\, \partial_j \,\mathcal{Q}\right )
= - \tilde k^2\mathcal{Q} \quad\mbox{where}\quad \tilde\nabla^2= a_0^2\,\nabla^2,~\tilde k=a_0\,k\,.
\label{helmholtz}
\end{equation}
Here $\gamma^{ij}\gamma_{jl}=\delta^i_l$ and $\tilde\nabla,\tilde k$ are dimensionless versions of the standard $\nabla, k$. For flat 3-space, it is standard practice to use the dimensionless scale factor $\tilde a\equiv a/a_0$ and to replace the coordinate $\chi$ in the metric by the dimensionful $r=a_0\chi$. 

Equation \eqref{helmholtz} is satisfied by separable functions $\mathcal{Q}(\chi,\theta, \phi) = R(\chi)Y_{lm}(\theta, \phi)$, where $Y_{lm}(\theta, \phi)$ are the spherical harmonics functions which satisfy 
\begin{equation}
 \frac{1}{\sin\theta}\,\partial_\theta \big ( \sin\theta\, \partial_\theta Y_{lm} \big ) + \frac{1}{\sin^2\theta}\partial_\phi^2 Y_{lm} = -l(l+1)Y_{lm}.
 \label{angular}
\end{equation}
Thus the radial part $R$ must satisfy 
\begin{equation}
 S_K^2\, \frac{{\rm d}^2 R}{\dif \chi^2} + 2C_K S_K\, \frac{\dif R}{\dif \chi} + \left [ \tilde k^2S_K^2 - l(l+1) \right ] R = 0.
 \label{radial}
\end{equation}

In the flat case $K = 0$, the solutions are spherical Bessel functions $j_l(\tilde k\chi)$ 
and the standard plane-wave basis is recovered via the identity
\begin{equation}
 {\rm e}^{{\rm i}\, \tilde{\bm k}\cdot\bm \chi} = \sum_{l=0}^\infty (2l+1){\rm i}^l\,j_l(\tilde k\chi)\, {L}_l(\hat{\tilde{\bm k}}\cdot\hat{\bm \chi}),
\end{equation}
where ${L}_l$ are Legendre polynomials. When $K\neq 0$, the general solution of \eqref{radial} for $R$ is
\begin{equation}
X_l^{(K)}(\nu, \chi) = (-K)^l\,{\big[S_K(\chi)\big]^l}\,\frac{{\rm d}^{l} X_0^{(K)}(\chi, \nu)}{\dif \big[C_K(\chi)\big]^l} \quad\mbox{where}\quad X_0^{(K)}(\nu,\chi) = \frac{\sin(\nu\chi)}{\nu\, S_K(\chi)}.
\label{rodrigues}
\end{equation}
Here $\nu$ is a generalised wave-number, related to the dimensionless Fourier wave-number in \eqref{helmholtz} by
\begin{equation}\label{nu}
\tilde k^2=\nu^2 - K\,, 
\end{equation}
where $\nu$ is an integer ($\geq2$) when $K=1$.

We can normalise the radial functions such that
\begin{equation}
4\pi\int_0^{\infty} \dif \chi\,S_K^2(\chi)\, \hat X_l^{(K)}(\nu, \chi)\,\hat X_l^{(K)}(\nu', \chi) =
\frac{2\pi^2}{\nu^{2}}\,
\left \{ 
\begin{array}{lc}
 \delta^{\rm D}(\nu-\nu') & \quad{\rm if}\quad K \le 0,\\
 & \\
 \delta_{\nu\nu'} & \quad{\rm if}\quad K = 1,
 \end{array}
 \right . 
\end{equation}
where
\begin{equation}
\hat X_l^{(K)}(\nu, \chi)= X_l^{(K)}(\nu, \chi)\,
\left\{
\begin{array}{lc}
 \big[(\nu^2-K)(\nu^2-2^2K)\cdots (\nu^2-l^2K) \big]^{1/2} & \quad{\rm if}\quad l \ge 1,\\
 & \\
 1 & \quad{\rm if}\quad l = 0.
 \end{array}
 \right .
\end{equation}
Then we have the additional orthogonality relations in Fourier space: 
\begin{eqnarray}
\int_0^{\infty}\dif \nu\, \frac{\nu^2}{2\pi^2} \hat X_l^{(K)}(\nu, \chi)\hat X_l^{(K)}(\nu, \chi') &=& \frac{\delta^{\rm D}(\chi-\chi')}{4\pi S_K^2(\chi)}\quad\mbox{when}\quad K \le 0\,,\\
\sum_{\nu=2}^{\infty} \frac{\nu^2}{2\pi^2} \hat X_l^{(K)}(\nu, \chi)\hat X_l^{(K)}(\nu, \chi') &=& \frac{\delta^{\rm D}(\chi-\chi')}{4\pi S_K^2(\chi)}
\quad\mbox{when}\quad K =1\,.
\label{orthos}
\end{eqnarray}

As a result, we can express the density contrast $\delta(\chi, \theta, \phi) \equiv \rho/\bar\rho -1$ of matter ($\delta=\delta_{\rm m}$) or galaxies ($\delta=\delta_{\rm g}$) as a Fourier series:
\begin{equation}
\delta (\chi, \theta, \phi) = 4\pi\int_0^\infty\dif \nu\, \nu^2\sum_{l=0}^{\infty}\sum_{m=-l}^l \delta_{lm}(\nu)\, \hat X_l^{(K)}(\nu,\chi)\,Y_{lm}(\theta, \phi),
\label{dchi}
\end{equation}
where $\delta_{lm}(\nu)$ are the Fourier coefficients. Note that when $K=1$ the continuous sum (integral) in \eqref{dchi} must be replaced by a discrete sum over $\nu$, as in \eqref{orthos}. 
Using the orthogonality relation: 
\begin{equation}
\delta_{lm}(\nu) = \frac{1}{2\pi^2} \int {\rm d}^2\Omega\, \dif\chi\, S_K^2(\chi)\, \delta(\chi, \theta, \phi) \hat X^{(K)}_l (\nu, \chi) Y_{lm}^*(\theta, \phi) \,. 
\label{dlm}
\end{equation}

Thus one can compute the cross-correlation between Fourier coefficients $\big\langle \delta_{lm}(\nu)\delta_{l'm'}^{*}(\nu')\big\rangle$ as an integral over all possible configurations of the $2$-point correlation $\xi$ between position ($\chi_1$, $\theta_1$, $\phi_1$) and ($\chi_2$, $\theta_2$ $\phi_2$). However, the cosmological principle implies that statistical properties of the universe should be invariant under translation and rotation so that the $2$-point correlation function should be a function only of the geodesic separation $\chi_{12}$ between point $1$ and $2$. Therefore one can define Fourier coefficients $S_{lm}(\nu)$ of the $2$-point correlation function such that 
\begin{equation}
S_{lm}(\nu) = \frac{1}{2\pi^2} \int {\rm d}^2\Omega\,\dif\chi_{12}\, S_K^2(\chi_{12}) \xi(\chi_{12}) \hat X^{{(K)}}_l (\nu, \chi_{12}) Y_{lm}^*(\theta, \phi) .
 \label{slm}
\end{equation}
Since the $2$-point correlation is a function only of $\chi_{12}$ and not of $\theta,\phi$ it follows that the only nonzero Fourier coefficient will be the monopole, which is the dimensionless power spectrum $\scal$: 
\begin{eqnarray}
\scal(\nu) &\equiv& S_{00}(\nu)=
\frac{1}{2\pi^2} \int {\rm d}^2\Omega\,\dif\chi_{12}\, S_K^2(\chi_{12}) \xi(\chi_{12}) \hat X^{(K)}_0 (\nu, \chi_{12}),
\label{scal}\\
\xi(\chi_{12}) &=& 4\pi\int \dif \nu\, \nu^2 \scal(\nu) \hat X^{(K)}_0(\nu, \chi_{12}).
\label{xi12}
\end{eqnarray}

\begin{figure}[!]
\begin{center}
\includegraphics[clip, trim = 3.4cm 7cm 2.7cm 7cm, width=0.6\linewidth]{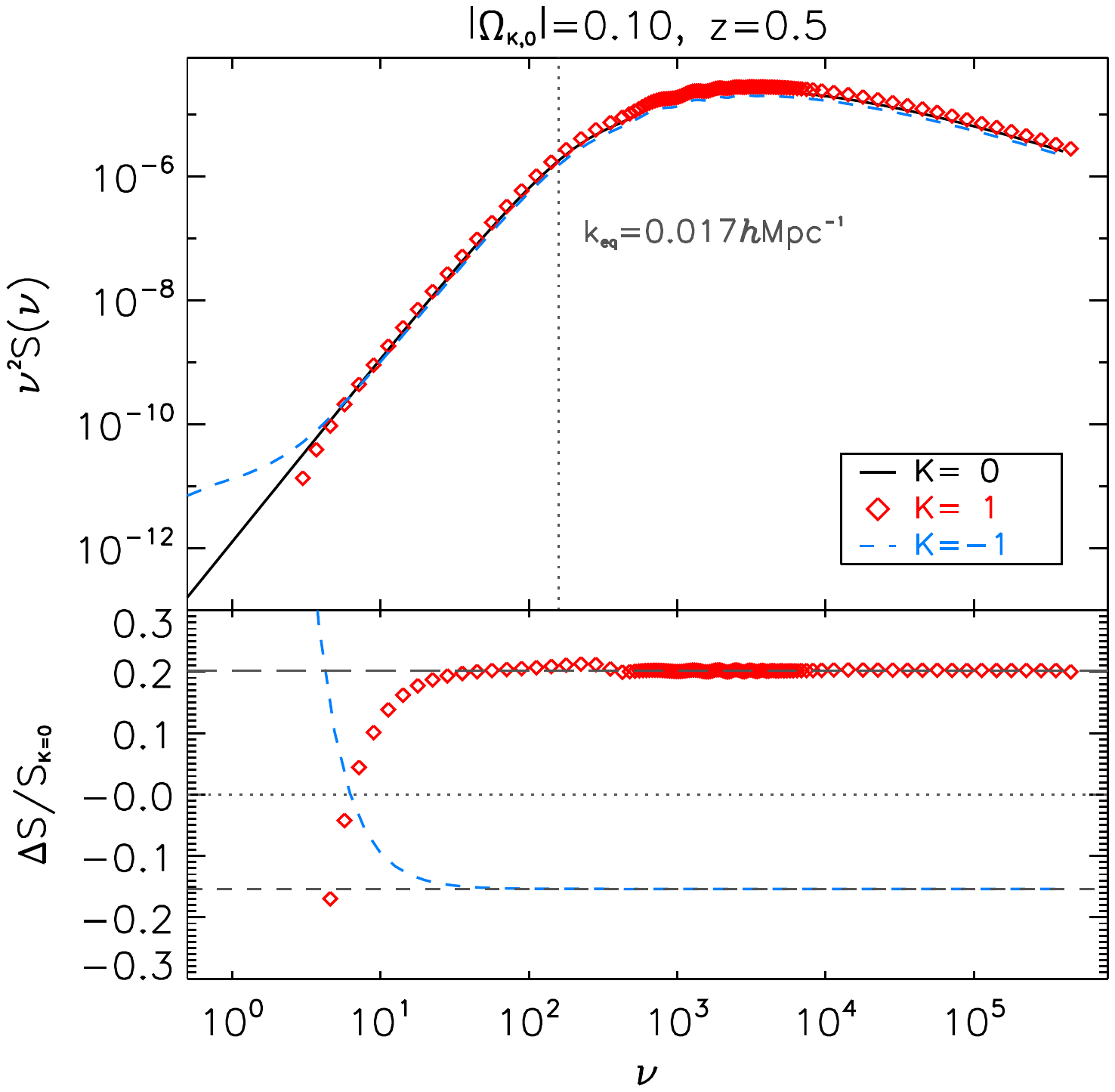}
\caption{{\it Top:} Power spectrum $\nu^2\scal(\nu)$ at $z=0.5$, for flat (black solid line), spherical (red diamonds) and hyperbolic (blue short dashed line) models, with $\Omega_{{\rm m},0}=0.32$ and $\Omega_{K,0}=0,\,\pm\, 0.1$. {\it Bottom:} Fractional difference relative to the flat model.} 
\label{fig:snu}
\end{center}
\end{figure}

Finally, the addition theorem 
\begin{equation}
\hat X^{(K)}_0(\nu, \chi_{12}) = 4\pi \sum_{q=0}^{\infty}\sum_{n=-q}^{q} \hat X^{(K)}_q(\nu, \chi_1) \hat X^{(K)}_q(\nu, \chi_2) Y_{qn}(\theta_1, \phi_1)Y_{qn}^*(\theta_2, \phi_2),
\label{add}
\end{equation}
leads to the fundamental equation that is implied by invariance under translation and rotation: 
\begin{equation}
\big\langle \delta_{lm}(\nu) \delta_{l'm'}^{*}(\nu')\big\rangle = \delta_{ll'}\,\delta_{mm'}\,\frac{\scal(\nu)}{\nu^2}\,
\left \{ 
\begin{array}{lc}
 \delta^{\rm D}(\nu-\nu') & \quad{\rm if}\quad {K} \le 0,\\
 & \\
 \delta_{\nu\nu'} & \quad{\rm if}\quad {K} = 1.
 \end{array}
 \right . 
\label{transopen}
\end{equation}
The relation to the usual matter power spectrum $P(k)$ is
\begin{equation}
 \nu\,\scal(\nu)= \frac{k}{a_0^2}\,P( k)
 \quad\mbox{where}\quad k= \frac{\tilde k}{a_0}=\frac{\sqrt{\nu^2-K}}{a_0} \quad\mbox{and}\quad \tilde k\,\chi = k\,r.
\end{equation} 
\autoref{fig:snu} shows $\scal(\nu)$ for $\Omega_{K,0}=0,\,\pm\, 0.1$. 

The two main effects of curvature on the power spectrum $\scal$ can be seen in \autoref{fig:snu}:
\begin{itemize}
\item An intrinsic contribution that changes the shape of the power spectrum on very large scales $\nu\lesssim 100$, close to the curvature scale $a_0$ given in \eqref{a0}. This follows from assuming nearly scale-invariant initial conditions for the gravitational potential that are the same regardless of $K$, leading to \cite{Matsubara:1999du}
\begin{equation}
\scal(\nu) = \frac{(\nu^2-4K)^2}{\nu (\nu^2-K)}\, A_{\rm s} \left ( \frac{k}{k_0} \right )^{n_{\rm s}-1} T^2(k), \end{equation}
where $T(k)$ is the matter transfer function, $n_s$ is the spectral index, $A_s$ is the scalar amplitude, $k_0$ is the pivot scale. The first factor accounts for curvature effects on the potential fluctuations and matter density, and is responsible for the curvature effects observed at large scales in \autoref{fig:snu}. 

\item The transfer function also induces additional curvature effects. These are in practice independent of the $\nu$ scale, and mostly due to the fact that evolving curvature changes the way perturbations grow in time, so that the growth factor $D(z)=\delta_{\rm m}(z,\bm r)/\delta_{\rm m}(0,\bm r)$ acquires a specific sensitivity to curvature. As a consequence, the overall amplitude of the power spectrum is modified for $\nu \gtrsim 100$. 
\end{itemize}

\subsection{Redshift-space distortions}\label{sec2.2}

The galaxy comoving number density contrast in real space is $\delta_{\rm g}=(n_{\rm g}-\bar n_{\rm g})/\bar n_{\rm g}$. 
In the linear regime, 
the redshift-space number density contrast is presented in \cite{Challinor:2011bk,Jeong:2011as,Bertacca:2012tp}:
\begin{eqnarray}\label{1a}
\delta^s_{\rm g}(z,\bm r)&=& b(z)\delta_{\rm m}(z,\bm r)-\frac{(1+z)}{H(z)}\,\frac{\partial}{\partial r}\big[\bm v(z,\bm r)\cdot \hat{\bm r} \big]
\notag \\&&{} 
- \frac{(1+z)}{H(z)} \alpha(z)\,\big[\bm v(z,\bm r)\cdot \hat{\bm r} \big]+\big[5s(z)-2 \big]\,\kappa(z,\bm r)
+ \delta_\Phi(z,\bm r) ,
\end{eqnarray}
where $b$ is the linear bias, $\bm v$ is the peculiar velocity, and $\kappa$ is the lensing convergence. The first line of \eqref{1a} is the standard `Newtonian' approximation with Kaiser redshift-space distortion (RSD). Typically a plane-parallel approximation is used, i.e. a global line of sight $\hat{\bm n}$ is assumed: $\hat{\bm r}~\to~\hat{\bm n}$. Here we do not impose the plane-parallel approximation, following \cite{Challinor:2011bk, Jeong:2011as, Bertacca:2012tp, DiDio:2016ykq, Matsubara:1999du}.

The second line of \eqref{1a} contains the relativistic corrections to RSD: a Doppler redshift effect; a contribution from lensing convergence, due to the modulation of solid angle and the effect of a magnitude limit $m_*$, where $s=\partial\log_{10} n_{\rm g}/\partial m_*$ is the magnification bias; and Sachs-Wolfe-type effects from the gravitational potential $\Phi$. The lensing contribution scales as the matter density contrast: $\kappa \sim \delta_{\rm m}$. It is negligible at low redshifts but can become important at high redshifts. For the current surveys that we consider, it can be safely neglected. The potential term $\delta_\Phi$ scales in Fourier space as $(H/k)^2\delta_{\rm m}$ by the Poisson equation, and we also neglect this contribution.

The Doppler term is less suppressed than $\delta_\Phi$, scaling as ${\rm i}\,(H/k)\delta_{\rm m}$ by the continuity equation. Although it is typically much smaller than the Kaiser term, it can be of the same order of magnitude as wide-angle effects in $\delta^s_{\rm g}$ \cite{Tansella:2017rpi}. However, its contribution to the 2-point correlations of $\delta^s_{\rm g}$ is more subtle. For a single tracer and for equal-redshift correlations, the Doppler term appears in the auto-correlations only as a square, i.e. its contribution is suppressed by $(H/k)^2$, like the leading contribution of $\delta_\Phi$. For the correlations of two tracers, or for single-tracer correlations at unequal redshifts, the dominant Doppler contribution scales as $(H/k)\delta_{\rm m}$ (see \cite{McDonald:2009ud, Bonvin:2013ogt, Bacon:2014uja, Lepori:2017twd}).

In other words, when considering the 2-point correlations of a single tracer in the same redshift bin, as we do here, the Doppler contribution is of the same order of magnitude as the potential contribution -- and it is therefore not consistent to introduce the Doppler term while neglecting $\delta_\Phi$. We thus neglect the Doppler term and consequently the second line of \eqref{1a}. 

The redshift-space correlation function can be expanded as \cite{Matsubara:1999du}
\begin{equation}
\xi^{s}_{\rm g}(\bm\chi_1,\bm \chi_2) = b_1b_2D_1D_2 \sum_{n,l}c_l^{(n)}(\chi_1,
\chi_2, \theta)\, \Xi_l^{(n)}(\chi),
\label{matsubara}
\end{equation}
and $b_i=b(\chi_i)$, $D_i=D(\chi_i)$. The angular separation between $\bm \chi_1$ and $\bm \chi_2$ is $\theta$, and $\chi$ is their comoving separation \cite{Matsubara:1999du}: 
\begin{eqnarray}
\cos\theta &=& \sin\theta_1\sin\theta_2\cos(\phi_2-\phi_1) + \cos\theta_1\cos\theta_2,\\ 
 C_K (\chi) &=& C_K(\chi_1)C_K(\chi_2) +K\, S_K(\chi_1)S_K(\chi_2)\cos\theta,
 \label{chicurved}
\end{eqnarray}
where $(\theta_i,\phi_i)$ describe the directions $\hat{\bm{\chi}}_i$. The functions $\Xi_l^{(n)}$ are determined by $\scal(\nu)$:
\begin{equation}
\Xi_l^{(n)}(\chi) = 4\pi(-1)^n
\int \dif \nu\, \frac{\nu^2}{(\nu^2-4K)^n}
\scal (\nu)X_l^{(K)}(\nu,\chi).
\label{transformln}
\end{equation}

In the sum \eqref{matsubara}, only terms with $0\le n\le 2$ and $0\le l\le 2n$ contribute. We can unify the three sets of coefficients given by \cite{Matsubara:1999du} into a single set:
\begin{eqnarray}
c_0^{(0)} & = & 1 + \frac{1}{3}(\beta_1+\beta_2) + \frac{1}{15}\beta_1\beta_2\left(1+2\cos^2{\vartheta}\right), \label{cnl} \\
c_0^{(1)} & = & \left [ \beta_1+\beta_2 + \frac{2}{15}\beta_1\beta_2\left ( 4+3\cos{\vartheta} \right )\right ] |K| - \frac{\beta_1\beta_2}{3}\alpha_1\alpha_2\cos{\vartheta}, \nonumber \\
c_1^{(1)} & = & \beta_1\alpha_1\cos\gamma_1+\beta_2\alpha_2\cos\gamma_2+\frac{\beta_1\beta_2}{15}\left [ \alpha_1\left ( \cos\gamma_1-2\cos\gamma_2\cos{\vartheta}\right ) + \alpha_2\left ( \cos\gamma_2 - 2\cos\gamma_1\cos{\vartheta} \right ) \right ], \nonumber \\
c_2^{(1)} & = & \frac{\beta_1}{3}\left ( 3\cos^2\gamma_1-1 \right ) + \frac{\beta_2}{3} \left ( 3\cos^2\gamma_2 - 1\right ) 
\notag\\&&{}
- \frac{\beta_1\beta_2}{21} \left [ {2} + {4}\cos^2{\vartheta} - 3\left(\cos^2\gamma_1+\cos^2\gamma_2\right) + 12\cos\gamma_1\cos\gamma_2\cos{\vartheta} \right ], \nonumber \\
c_0^{(2)} & = & \beta_1\beta_2\left ( 1-\alpha_1\alpha_2 \right )|K|, \nonumber \\
c_1^{(2)} & = & \beta_1\beta_2\left ( \alpha_1\cos\gamma_1 + \alpha_2\cos\gamma_2\right )|K|, \nonumber \\
c_2^{(2)} & = & \frac{\beta_1\beta_2}{21} \left [ 6\cos^2{\vartheta} - 4 + 27\left( \cos^2\gamma_1 + \cos^2\gamma_2 \right)+60\cos\gamma_1\cos\gamma_2\cos{\vartheta} \right ] |K| 
\notag\\&&{}
+ \frac{\beta_1\beta_2\alpha_1\alpha_2}{3}\left( 3\cos\gamma_1\cos\gamma_2 + \cos{\vartheta} \right), \nonumber \\
c_3^{(2)} & = & \!\frac{\beta_1\beta_2}{5}\!\left [ \alpha_1 \!\left( 5\cos\gamma_1\cos^2\gamma_2-\cos\gamma_1 + 2\cos\gamma_2\cos{\vartheta}\right) \!
+ \alpha_2\! \left( 5\cos\gamma_2\cos^2\gamma_1 - \cos\gamma_2 + 2\cos\gamma_1\cos{\vartheta} \right)\! \right]\!\!, \nonumber \\
c_4^{(2)} & = &\! \frac{\beta_1\beta_2}{35}\!\left [ 1 + 2\cos^2{\vartheta} - 5\left(\cos^2\gamma_1+\cos^2\gamma_2\right) + 20\cos\gamma_1\cos\gamma_2\cos{\vartheta} + 35\cos^2\gamma_1\cos^2\gamma_2 \right ] .  \nonumber          
\end{eqnarray}
Here $\beta_i = {f_i}/{b_i}$ and $f=-\dif\,\ln D/\dif\,\ln(1+z)$ is the growth rate.

The $\gamma_i$ are the angles between $\bm \chi_i$ and the separation vector $\bm \chi_{12}=\bm{\chi}_1-\bm{\chi}_2 $, while ${\vartheta}$ is defined as \cite{Matsubara:1999du}
\begin{eqnarray}
\cos{\vartheta} &=& \frac{\sin\gamma_1\sin\gamma_2}{C_{12}} - \cos\gamma_1\cos\gamma_2 = \frac{C_1C_2\cos\theta + K S_1S_2}{C_1C_2 + K S_1S_2\,\cos\theta},
\label{cothetat}
\end{eqnarray}
where $S_i= S_K(\chi_i)$, $C_i= C_K(\chi_i)$, $S_{12}= S_K(\chi)$ and $C_{12}= C_K(\chi)$. Note that $\vartheta=\theta$ for $K=0$. We also have
\begin{eqnarray}
\cos \gamma_1S_{12} & = & S_1C_2 - C_1S_2\cos\theta ,\\
\cos \gamma_2S_{12} & = & C_1S_2 - S_1C_2\cos\theta .
\end{eqnarray}

As pointed out above, we set $\alpha_i=0$ since we consider correlations of a single tracer in the same redshift bin. (Note that the $\alpha_i$ given in \cite{Matsubara:1999du} omit some contributions and are corrected by the expressions in \cite{DiDio:2016ykq}.)

In the absence of RSD ($\beta_i=0$), the real space correlation function reduces to 
\begin{equation}
\frac{\xi_{\rm g}(\chi)}{b_1b_2D_1D_2}= 4\pi\int_0^\infty \dif\nu\, \nu^2\scal(\nu)X_0^{(K)}(\nu,\chi)
= \frac{4\pi}{S_K(\chi)} \int_0^\infty \dif\nu\, \nu\,\scal(\nu)\sin(\nu\chi).
\label{xinu}
\end{equation}
From this we can express the integral as an inverse Fourier transform (useful for practical calculations using \verb!FFT! or \verb!FFTLog! algorithms):
\begin{equation}\label{fft}
\frac{\xi_{\rm g}(\chi)}{b_1b_2D_1D_2} = \frac{2\pi}{S_K(\chi)} \int_{-\infty}^{\infty}\dif\nu\, \nu\,\scal(\nu){\rm e}^{{\rm i}\,\nu \chi} = \frac{2\pi}{S_K(\chi)} \fcal^{-1}[\nu\, \scal(\nu)].
\end{equation}
Note that $\nu\scal(\nu)$ and $S_K (\chi)\xi(\chi)/(2\pi b_1b_2D_1D_2)$ form a Fourier pair. Equation \eqref{fft} can be expressed as a Hankel transform
\begin{equation}
\frac{\xi_{\rm g}(\chi)}{b_1b_2D_1D_2} = \frac{4\pi\chi}{S_K(\chi)} \int_0^\infty \dif\nu\, \nu^2\scal(\nu)j_0(\nu\chi).
\label{hankel}
\end{equation}

Based on \eqref{xinu} or \eqref{hankel}, we can decompose the curvature effects on the real-space correlation function into three contributions: (a)~as noted before an intrinsic effect of curvature on the shape of the power spectrum $\scal(\nu)$ and (b)~an effect through the growth factor $D$ (which is scale independent); (c)~a scale-dependent effect when $K\neq 0$ in the relation between the power spectrum and the correlation function, via the factor $S_K(\chi)$ in \eqref{xinu}. 

In redshift space, curvature and wide-angle effects appear on a range of scales, leading to a more complicated picture than in real space. We start with a simplified plane-parallel model:

(a)~assuming the distant-observer limit, $\chi\ll \chi_1,\,\chi_2$ with $\gamma_i$ fixed; 

(b)~neglecting evolution effects, i.e., $D_1=D_2=D$, $b_1=b_2=b$ and $f_1=f_2=f$.\\ 
Note that (a) implies that curvature can be neglected in this limit, since $\chi$ is much less than the curvature scale; (a)~also implies that $\theta,{\vartheta} \to 0$ and $\gamma_2\to\pi-\gamma_1$. Then only 3 of the $c_l^{(n)}$ survive and \eqref{matsubara} reduces to the Kaiser plane-parallel result,
\begin{equation}
\xi_{\rm pp}(\chi, \mu) = \xi^{(0)}_{\rm pp}(\chi) + \xi^{(2)}_{\rm pp}(\chi)L_2(\mu) + \xi^{(4)}_{\rm pp}(\chi)L_4(\mu)\quad\mbox{where}\quad {\mu= \cos\gamma_1}.
\label{xipp}
\end{equation}
The multipoles are\footnote{{Note that for the cross-correlation of two different tracers, there is a nonzero dipole and octupole in the plane-parallel limit \cite{McDonald:2009ud,Bonvin:2013ogt,Lepori:2017twd}.}}
\begin{eqnarray}
\xi^{(0)}_{\rm pp} & = & b^2D^2\left ( 1 + {2\over3}\beta +{1\over5} \beta^2 ) \right ) \Xi_0^{(0)}, \label{monop} \\
\xi^{(2)}_{\rm pp} & = & b^2D^2 \left ( {4\over3}\beta + {4\over7}\beta^2 \right ) \Xi_2^{(1)}, \label{quadri} \\
\xi^{(4)}_{\rm pp} & = & {8\over 35}b^2D^2 \beta^2\, \Xi_4^{(2)}. \label{hexad}
\end{eqnarray}
The spatial curvature enters via the real-space $l$-multipoles $\Xi_l^{(n)}$, given by \eqref{transformln}. 

\subsection{Wide-angle effects}\label{sec2.3}

We now consider the redshift-space correlation function with wide-angle effects, and compare it with the plane-parallel limit. Using \eqref{matsubara} we can predict the redshift space $2$-point correlation function (taking into account both curvature and wide-angle effects). First we choose an observer line-of-sight direction. Different choices lead to small differences (see e.g. \cite{Reimberg:2015jma,Castorina:2017inr}) which can be neglected for the surveys that we consider. We choose $\hat{\bm{x}}_1$ as line-of-sight direction ($\bm{x}_1 =a_0\bm{\chi}_1$). We specify the redshift $z_1$ at the end-point of $\bm{x}_1$, the comoving separation $r $ between the end-points $\bm{x}_1$ and $\bm{x}_2$, and the angle $\gamma_1$ between the line-of-sight and the geodesic between end-points $\bm{x}_1$ and $\bm{x}_2$. This can be done for several values of the comoving separation $r$ and appropriate values of $\mu$, so that we can compute $\xi^s_{\rm g}(r,\mu)$. 

The multipoles are then given by 
\begin{equation}\label{2pcfmp}
\xi^{(l)}(r)=\frac{1}{2}(2l+1)\int_{-1}^1\dif \mu\, L_l(\mu)\,\xi^s_{\rm g}(r,\mu)\,.
\end{equation}
Beyond the plane-parallel limit, wide-angle corrections arise in the even multipoles -- and they source a dipole and octupole.

\begin{figure}
\centering
\includegraphics[clip, trim = 1.7cm 19.4cm 0.7cm 2cm, width=0.95\linewidth]{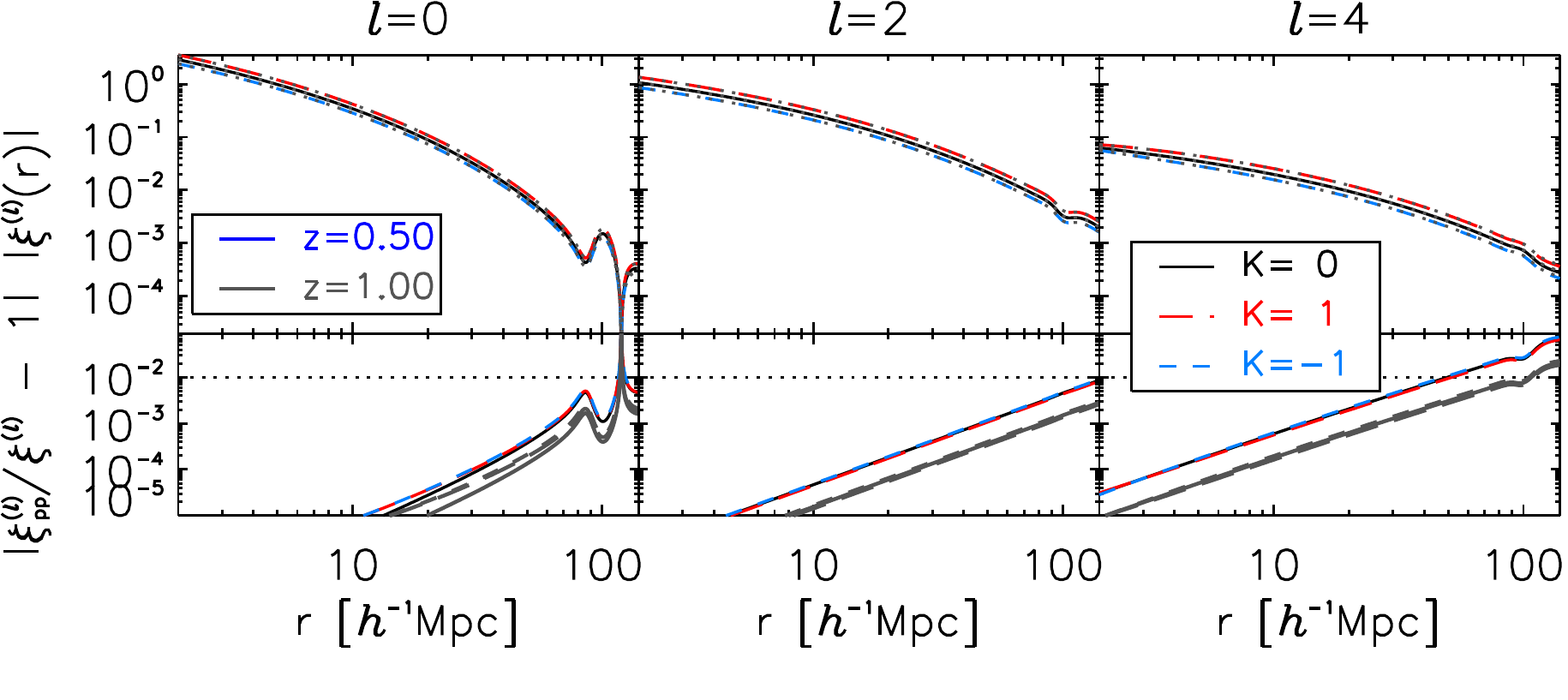}
\caption{{\it Top:} $2$-point correlation function monopole (left), quadrupole (middle) and hexadecapole (right) for flat (black), spherical (red) and hyperbolic (blue) spaces, at $z=0.5$. The plane-parallel (Kaiser) limit is shown in dotted curves. {\it Bottom:} Fractional difference compared to the Kaiser limit at $z=0.5$ (same colour code as top panel) and at $z=1$ (grey thick lines). Dotted line is a 1\% relative difference.} 
\label{fig:even}
\end{figure}

\autoref{fig:even} shows the redshift-space even multipoles in the three curvature cases, at two redshifts. The upper panels display the general case and the plane-parallel limit [\eqref{monop}--\eqref{hexad}]. The lower panels give the fractional difference. On scales below $70h^{-1}$Mpc, wide-angle effects change the monopole by less than $0.1$\%. However, around the BAO scale the wide-angle effect is around $0.4$\% and can reach a maximum of $0.6$\% at redshift $z=0.5$. By contrast, at redshift $z=1$ the effect on the BAO scale remains lower than $0.2$\%. While the quadrupole is affected at the percent level at $150h^{-1}$Mpc at redshift $0.5$, the hexadecapole shows the largest deviation of about $1$\% at $r\sim 50h^{-1}$Mpc for $z=0.5$ and $1$\% at $r\sim 100h^{-1}$Mpc for $z=1$. In general, it is interesting that wide-angle effects are lowered by a factor of $2$ on all the multipoles when redshift increases from $0.5$ to $1$. Whether these effects are relevant for BAO measurements is beyond the scope of the present paper, and this would need to be contrasted with typical error measurements in a given galaxy survey. 

\begin{figure}[!]
\begin{center}
\includegraphics[clip, trim = 1.7cm 19.5cm 3.1cm 0.2cm, width=0.75\linewidth]{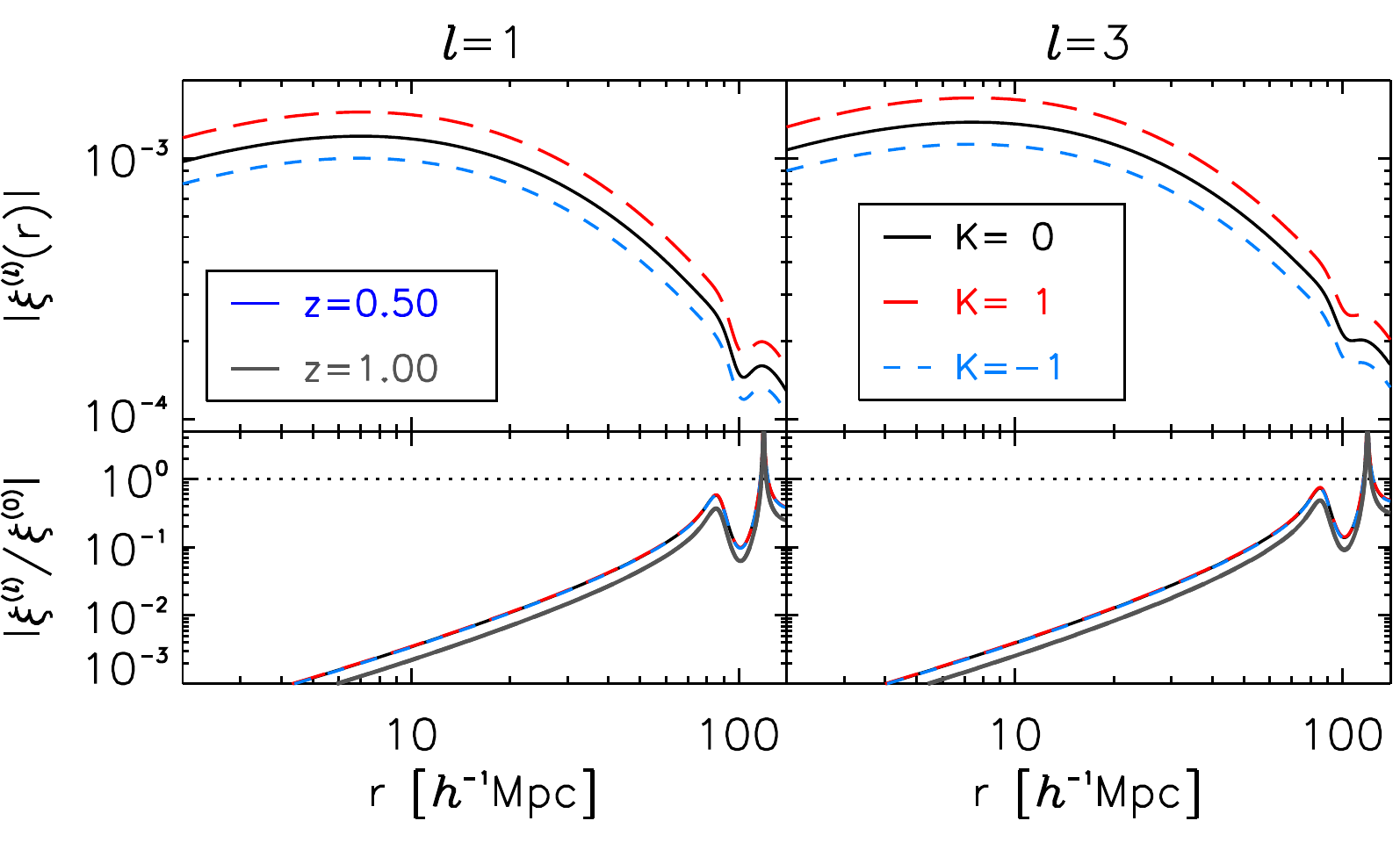}
\caption{As in \autoref{fig:even}, but for odd multipoles.
{\it Top:} Odd multipoles at $z=0.5$. {\it Bottom:} Ratio with the monopole at $z=0.5$ (in colours) and at $z=1$ (in grey).} 
\label{fig:odd}
\end{center}
\end{figure}

The characteristic imprint of wide-angle effects is located in the odd multipoles, as illustrated in \autoref{fig:odd}. These vanish (by definition) in the plane-parallel approximation. The nonzero dipole (left pannel) and octupole (right panel) are respectively negative and positive. In order to quantify their importance we compute their ratio with the monopole (which is driving the signal) in the bottom panel of \autoref{fig:odd}. It appears that they start being relevant ({\it i.e.} a factor of $10$ smaller than the monopole) around the BAO scale at the two considered redshifts $0.5$ and $1$. Finally, comparing the two panels of \autoref{fig:odd}, the contributions of the dipole and octupole are roughly the same. We will investigate the possibility of detecting the odd multipoles in a future work. 

\begin{figure}[!]
\begin{center}
\hspace*{-0.7cm}
\vspace*{-13.0cm}
\includegraphics[scale=0.75]{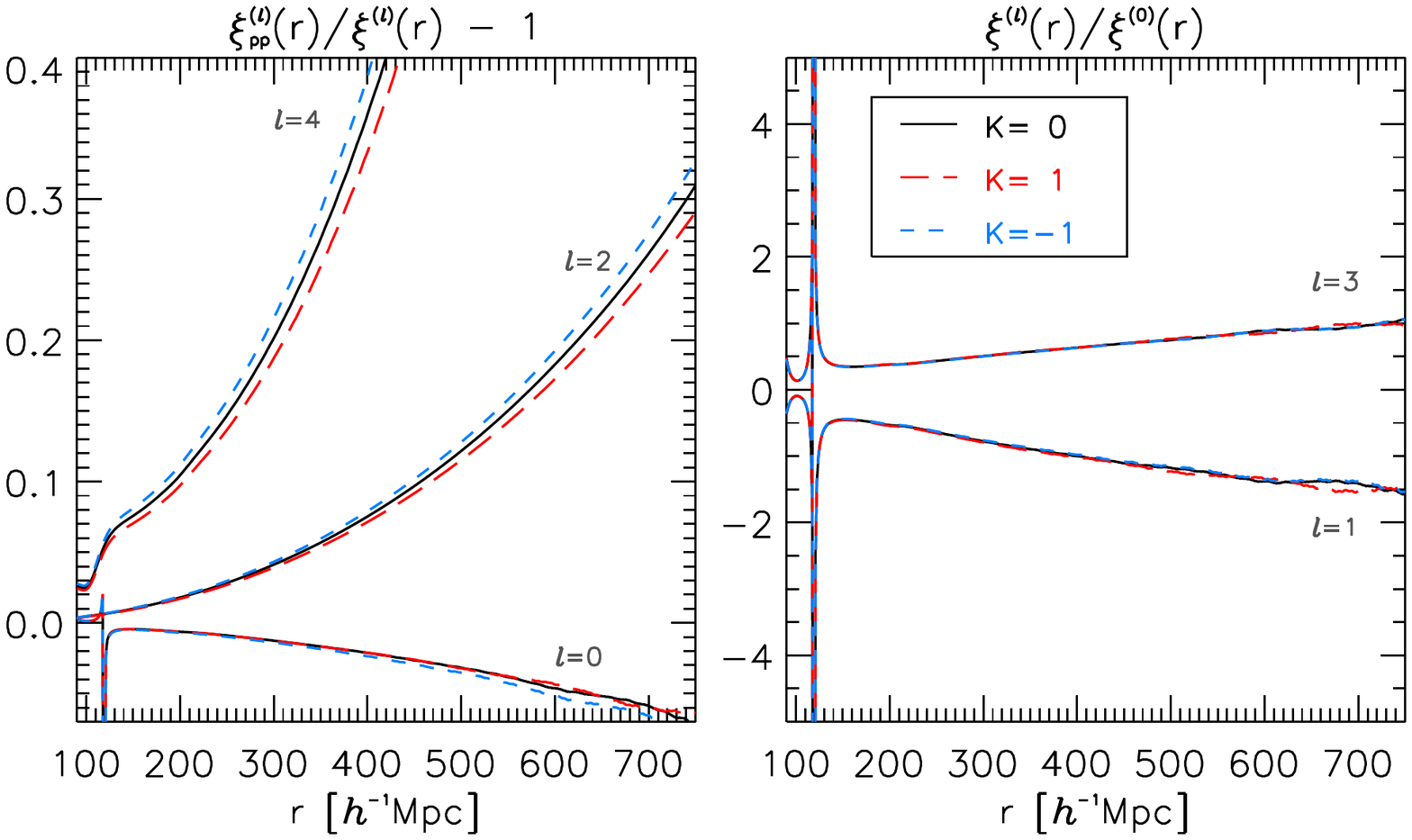}
\caption{{\it Left:} Fractional difference with the plane-parallel limit for even multipoles at $z=0.5$. {\it Right:} Ratio of odd multipoles to monopole at $z=0.5$. } 
\label{fig:rmax}
\end{center}
\end{figure}

For completeness, we also studied wide-angle effects on the even multipoles for an extended scale range $90<r<700h^{-1}$Mpc, which reaches $\sim$10\% of the curvature scale for $|\Omega_K|=0.1$. \autoref{fig:rmax} (left panel) shows that the hexadecapole is the multipole most sensitive to wide-angle effects. Indeed, the plane-parallel approximation over-predict the hexadecapole by $10$\% at scales around $200h^{-1}$Mpc. On the contrary, the quadrupole and monopole are less affected: the quadrupole is $10$\% over-estimated at $450h^{-1}$Mpc, while the monopole is under-estimated by less than $10$\% at $750h^{-1}$Mpc. It is also apparent that wide-angle effects are only marginally dependent on the curvature. In addition, on large scales we see in the right panel of \autoref{fig:rmax} that wide-angle effects modulate the ratios of dipole and octupole to monopole in the same way. It is interesting that odd multipoles reach $50$\% of the monopole at a scale $\sim$150$h^{-1}$Mpc, risding to $100$\% at $\sim$400$h^{-1}$Mpc. We also note that there is no dependence at all on curvature in the ratios of odd multipoles to the monopole. 

To conclude the present section, we see that wide-angle effects can alter the $2$-point correlation function monopole by up to $0.5$\% on scales above $80h^{-1}$Mpc at low redshift ($z\sim 0.5$). Since we will consider measurements of the Clustering Ratio which involve scales of about this range, we need to assess how much wide-angle effects can alter it. 

\subsection{Clustering Ratio}\label{sec:cr}

We have shown how wide-angle and curvature effects have a non-trivial impact on the {galaxy} $2$-point correlation function in redshift space. We now apply the analysis to study how they affect a different, though related, clustering statistic, namely the Clustering Ratio {of matter}, defined as \cite{Bel:2013csa,Bel:2013ksa,Zennaro:2017qnp}: 
\begin{equation}
 \eta_R(r) \equiv \frac{\xi_R^{(0)}(r)}{\sigma_R^2},
 \label{etaratio}
\end{equation} 
where $R$ is the smoothing scale for the matter density contrast:
\begin{equation}
\delta_R (\bm x) = \int {\rm d}^3\bm{x}'\, W_{R}( \bm x-\bm{x}' ) \, {\delta_{\rm m}}(\bm{x}').
\label{deltar}
\end{equation}

Here $W_{R}$ is a spherical top-hat filter and the smoothed $2$-point correlation function is 
\begin{equation}
\xi_R(\bm x_1, \bm x_2) = \int {\rm d}^3\bm y_1{\rm d}^3\bm y_2\, W_R(\bm y_2 - \bm x_2)\,W_R( \bm y_1-\bm x_1 )\,\xi_{\rm m}(\bm y_1, \bm y_2) .
 \label{xismooth}
\end{equation}

Without additional assumptions on the properties of the $2$-point correlation in redshift space, it is not possible to further simplify the integral \eqref{xismooth}. Therefore, in order to evaluate it, we use a Monte Carlo integration method: we draw a uniform distribution of $\bm y_1$ and $\bm y_2$ within their respective integration domain. Then at fixed $\bm x_1$ it is possible to evaluate $\xi_R(r,\mu)$, where $r$ is the comoving distance between $\bm x_1$ and $\bm x_2$ and {$\mu=\cos\gamma_1$ (see \autoref{sec2.3})}. We can finally estimate the amplitude of the multipoles $\xi_R^{(l)}(r)$, up to order $4$, from \eqref{2pcfmp}, using a Gauss-Legendre quadrature. As a result, if $\xi_R(r,\mu)$ is well described by a polynomial of order $9$ (its first nine multipoles), then the function that needs to be integrated is a polynomial of order $13$ (because we will focus on multipoles of order $\leq 4$). Thus the result of the quadrature is exact if we choose a Gauss-Legendre quadrature of order $7$ ($N = (n+1)/2$). In this case, the only source of error in the integration process will be the Monte Carlo error. This is why we evaluate the integral $80$ times to estimate the error on our Monte Carlo integration.

\begin{figure}[!]
\centering
\includegraphics[clip, trim = 1.2cm 17.4cm 1.5cm 1cm, width=0.95\linewidth]{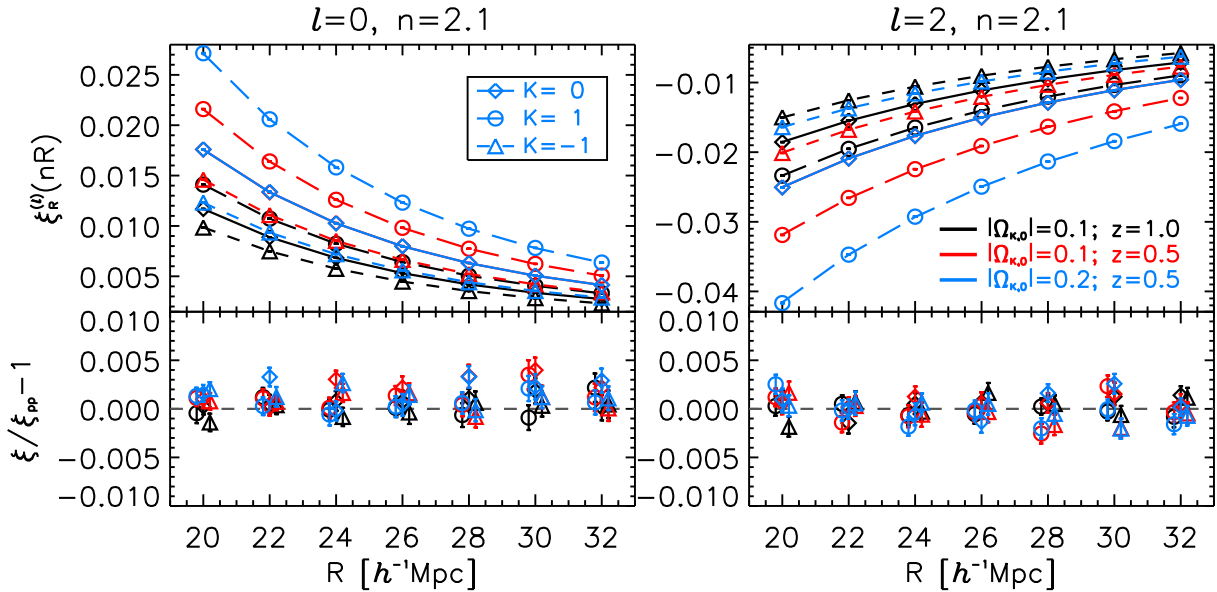}
\caption{{\it Top:} Smoothed monopole $\xi^{(0)}_R(nR)$ (left) and quadrupole $\xi^{(2)}_R(nR)$ (right), at the correlation length $nR=2.1\,R$, for flat (diamonds), spherical (circles) and hyperbolic (triangles) spaces, in 3 different models. Lines over-plotted show the plane-parallel limit. {\it Bottom:} Fractional difference relative to plane-parallel limit.} 
\label{fig:eveneta}
\end{figure}

On the other hand, if the plane-parallel approximation is adopted, the integral \eqref{xismooth} can be simplified. This leads to the same formal expression as for the $2$-point correlation function, given in \eqref{monop}--\eqref{hexad}: 
\begin{eqnarray}
\xi_{{\rm pp}R}^{(0)}(r) & = & b^2D^2\left ( 1 + {2\over3}\beta + {1\over5}\beta^2 ) \right ) \Xi_R^{(0,0)}(\chi), \\
\xi_{{\rm pp}R}^{(2)}(r) & = & b^2D^2 \left ( {4\over3}\beta + {4\over7}\beta^2 \right ) \Xi_R^{(1,2)}(\chi), \\
\xi_{{\rm pp}R}^{(4)}(r) & = & {8\over35}b^2D^2 \beta^2\, \Xi_R^{(2,4)}(\chi),
\end{eqnarray}
where 
\begin{equation}
   \Xi_R^{(n,l)}( \chi ) \equiv 4\pi (-1)^n\int \frac{\dif \nu}{(\nu^2-4K)^n}\, \nu^2\scal (\nu)\left [ W^{(K)}(\nu, R/a_0) \right ]^2X_l^{(K)}(\nu,\chi).
   \label{transformlnr}
\end{equation}
The function $W^{(K)}$ can be computed from \eqref{deltar}:
\begin{equation}
W^{(K)}( \nu,\chi ) = \frac{4\pi}{V^{(K)}(\chi)}\,\frac{1}{\nu(\nu^2-K)}\big [ C_K(\chi)\sin(\nu\chi) - \nu S_K(\chi)\cos(\nu\chi)\big],
\end{equation}
where
\begin{equation}
    V^{(K)}(\chi) = K\pi a_0^3\left [ 2\chi - S_K(2\chi) \right ],   
\end{equation}
is the volume of the sphere of radius $R = a_0\chi$. Note that in the flat case $V^{(0)}(\chi) = {4\pi}(a_0\chi)^3/3$, so that $W^{(0)}(\nu, \chi) = {3}j_1(\nu\chi)/({\nu\chi})$, which corresponds to the usual expression obtained when performing the (flat) Fourier transform of a top-hat spherical filter.

In the following, we compare in a given cosmological model the result obtained with and without assuming the plane-parallel limit. It is fundamental to make this comparison because the implementation of the plane-parallel approximation for computing the smoothed $2$-point correlation function is computationally more efficient.

In \autoref{fig:eveneta} we show that the plane-parallel approximation works extremely well as far as estimations of the monopole and quadrupole terms are concerned. The approximation and the exact numerical calculations agree to better than $0.5$\%. This result holds even when considering extreme curvature values as high as $|\Omega_{K,0}|=0.2$ and a wide redshift range $0.5$ to $1$. 
 
Although not shown here, we verified that the hexadecapoles agree at the level of $5$\% (without finding any significant deviation, since the error is mostly due to our Monte Carlo integration technique). Also note that we can safely assume that the variance $\sigma_R^2$ is correctly recovered under the plane-parallel approximation, since its estimate involves scales that are smaller than those considered when measuring the $2$-point correlation function. Only the monopole and the variance are needed to estimate the Clustering Ratio so we conclude that the plane-parallel approximation can be safely applied. 
 
\begin{figure}[!]
\centering
\includegraphics[clip, trim = 1.8cm 17.4cm 2cm 1cm, width=0.92\linewidth]{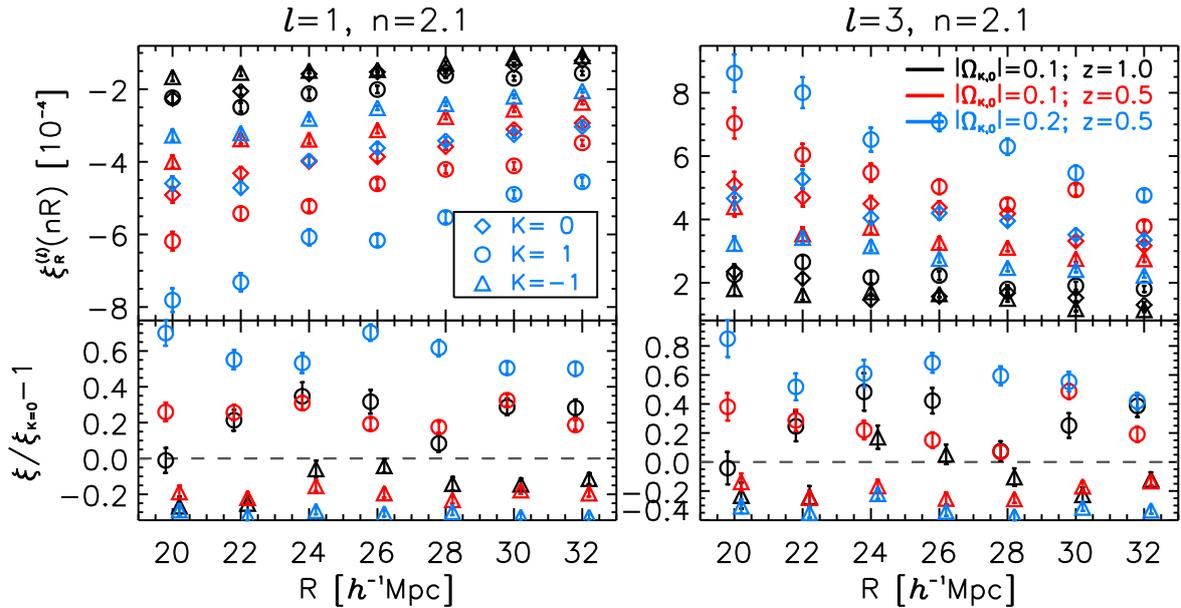}
\caption{\small Odd multipoles $\xi_R^{(l)}(nR)$ of the smoothed $2$-point correlation function at the correlation length $nR=2.1\,R$, for various curvature and redshift ranges. {\it Bottom:} Fractional difference relative to the $K=0$ case in each cosmology.} \label{fig:oddeta}
\end{figure}

For the sake of completeness, we also predict the amplitude of the lower odd multipoles, shown in \autoref{fig:oddeta}. Interestingly, significant contributions of RSD to the dipole and octupole are detected. The lower panels show the fractional difference relative to a flat-space model. This fractional difference depends on the amount of curvature but not on redshift. Whether or not this large-scale effect can be detected in redshift surveys (current or future) is left for a future analysis.

In summary, on the scales we consider, wide-angle effects or curvature effects are expected to impact negligibly on the estimation of the Clustering Ratio $\eta_R$. Thus one can safely apply the Kaiser model, confirming that the real-space matter Clustering Ratio can be inferred from galaxies as the ratio between the smoothed $2$-point correlation monopole $\xi_R^{(0)}$ and the variance $\sigma_R^2$ on the corresponding scale $R$. As a result, {\em the galaxy Clustering Ratio maintains its characteristic property of being insensitive to both galaxy bias and large-scale RSD,} thus remaining informative about the underlying matter field. 

\section{Methodology}\label{sec:data}

We aim to extract cosmological information in the most background-independent way possible, i.e. without relying on geometric probes, or probes that are primarily sensitive to the expansion rate. To this end, we focus on the amplitude and shape of the matter power spectrum, using the two observables: the Clustering Ratio $\eta_R$ and the RSD parameter $f \sigma_8$. The Clustering Ratio is sensitive to the shape of the power spectrum (see \autoref{sec:cr}), while, in addition, the RSD parameter is sensitive to the overall normalisation of the power spectrum via $ \sigma_8$. 

Then we complement our analysis, in order to strengthen our conclusion, by also using the information contained in additional datasets that are mainly sensitive to the geometry and kinematics (SNIa and BAO) or the physics of the early universe (CMB). 

\subsection{Probes and likelihoods}\label{sec:likelihoods}

Here we present the main probes used in the analysis, the chosen data and likelihoods, and the correcting schemes implemented to account for curvature effects. The likelihoods are in the ECLAIR suite \cite{Ilic:2020onu} and available online\footnote{\url{https://github.com/s-ilic/ECLAIR}, \url{https://github.com/louisperenon/ECLAIR_likelihoods}.}. 
\\

\noindent\textbullet~ {\bf Redshift-space distortions}\\ 
A large compilation of independent measurements of $f \sigma_8$ is given in \cite{Perenon:2019dpc} and summarised in \autoref{tab:rsd_data}. This data set is identified as `RSD' in figures. Note that some of the measurements have been obtained by marginalising over the AP parameters and some have not but we treat them in the same way. The reason is that the uncertainties on measurements without marginalisation are larger than the errors on the most recent  BOSS measurements which are done with AP marginalisation. Therefore we do not expect the results to change.

Since the RSD parameters have been estimated in some chosen fiducial cosmology (often different from survey to survey), we face the issue of forecasting their amplitude in any given cosmological model. Instead of following this route, we adopt the standard practice of prescribing a transformation rule that maps the $f\sigma_8$ predicted in a model, into $\widetilde {f\sigma}_8$ in a fiducial model. There is no consensus in the literature on how this mapping should be achieved. It is typical to prescribe a global rescaling:

\begin{equation}\label{qth}
\widetilde {f\sigma}_8 = q\, f\sigma_8\,. 
\end{equation}
The ansatz 
\begin{eqnarray}
q= \frac{\alpha_\parallel}{\alpha_\perp} \,,
\end{eqnarray}
is proposed by \cite{Macaulay:2013swa,Kazantzidis:2018rnb}, where the re-scaling parameters are 
\begin{eqnarray}
 \alpha_\parallel = \frac{\tilde E(\bar z)}{E(\bar z)}\,, \qquad
 \alpha_\perp = \frac{D_A(\bar z)}{\tilde D_A(\bar z)}\,.
\end{eqnarray}

Here $E\equiv H/H_0$ is the normalised expansion rate, $D_A$ is the angular diameter distance, $\bar z$ is the effective redshift of the measurement, and a tilde indicates evaluation in the fiducial cosmology.

Without physical justification of this mapping, and since previously developed schemes all apply to the spatially flat case, we have reconsidered the whole issue in \autoref{apeffect}. We find a theoretically motivated transformation: 
\begin{equation}\label{qus}
  q \simeq \frac{5}{7}+\frac{2}{7}\, \frac{\alpha_\parallel}{\alpha_\perp}\,,
\end{equation}
and we give the steps to obtain it, the assumptions on which it is based and simulations to test its accuracy. 

\begin{table}[!]
\small
\centering
\begin{tabular}{c c c c}
\hline
Data set & $z$ & $\fsig$ & Reference \\
\hline\hline
2MTF         & 0.001 & 0.505 $\pm$ 0.085 & \cite{Howlett:2017asq}\\
6dFGS+SNIa   & 0.02 & 0.428 $\pm$ 0.0465 & \cite{Huterer:2016uyq}\\
IRAS+SNIa    & 0.02 & 0.398 $\pm$ 0.065 & \cite{Hudson:2012gt,Turnbull:2011ty}\\ 
2MASS        & 0.02 & 0.314 $\pm$ 0.048 & \cite{Hudson:2012gt,Davis:2010sw}\\ 
SDSS         & 0.10 & 0.376 $\pm$ 0.038 & \cite{Shi:2017qpr}\\ 
SDSS-MGS     & 0.15 & 0.490 $\pm$ 0.145 & \cite{Howlett:2014opa}\\ 
2dFGRS       & 0.17 & 0.510 $\pm$ 0.060 & \cite{Song:2008qt}\\ 
GAMA         & 0.18 & 0.360 $\pm$ 0.090 & \cite{Blake:2013nif}\\ 
GAMA         & 0.38 & 0.440 $\pm$ 0.060 &\cite{Blake:2013nif}\\ 
SDSS-LRG-200 & 0.25 & 0.3512 $\pm$ 0.0583 & \cite{Samushia:2011cs}\\ 
SDSS-LRG-200 & 0.37 & 0.4602 $\pm$ 0.0378 & \cite{Samushia:2011cs}\\ 
BOSS DR12    & 0.31 & 0.469 $\pm$ 0.098 &  \cite{Wang:2017wia}\\
BOSS DR12    & 0.36 & 0.474 $\pm$ 0.097 &  \cite{Wang:2017wia}\\
BOSS DR12    & 0.40 & 0.473 $\pm$ 0.086 &  \cite{Wang:2017wia}\\
BOSS DR12    & 0.44 & 0.481 $\pm$ 0.076 &  \cite{Wang:2017wia}\\
BOSS DR12    & 0.48 & 0.482 $\pm$ 0.067 &  \cite{Wang:2017wia}\\
BOSS DR12    & 0.52 & 0.488 $\pm$ 0.065 &  \cite{Wang:2017wia}\\
BOSS DR12    & 0.56 & 0.482 $\pm$ 0.067 &  \cite{Wang:2017wia}\\
BOSS DR12    & 0.59  & 0.481 $\pm$ 0.066 &  \cite{Wang:2017wia}\\
BOSS DR12    & 0.64  & 0.486 $\pm$ 0.070 &  \cite{Wang:2017wia}\\
WiggleZ      & 0.44  & 0.413 $\pm$ 0.080 & \cite{Blake:2012pj}\\ 
WiggleZ      & 0.60  & 0.390 $\pm$ 0.063 & \cite{Blake:2012pj}\\ 
WiggleZ      & 0.73  & 0.437 $\pm$ 0.072 & \cite{Blake:2012pj}\\ 
Vipers PDR-2 & 0.60  & 0.550 $\pm$ 0.120 & \cite{delaTorre:2016rxm,Pezzotta:2016gbo}\\ 
Vipers PDR-2 & 0.86  & 0.400 $\pm$ 0.110 & \cite{delaTorre:2016rxm,Pezzotta:2016gbo}\\ 
FastSound    & 1.40  & 0.482 $\pm$ 0.116 & \cite{Okumura:2015lvp}\\ 
SDSS-IV      & 0.978 & 0.379 $\pm$ 0.176 & \cite{Zhao:2018gvb}\\
SDSS-IV      & 1.23  & 0.385 $\pm$ 0.099 & \cite{Zhao:2018gvb}\\
SDSS-IV      & 1.526 & 0.342 $\pm$ 0.070 & \cite{Zhao:2018gvb}\\
SDSS-IV      & 1.944 & 0.364 $\pm$ 0.106 & \cite{Zhao:2018gvb}\\
\hline
\end{tabular} 
\caption{RSD data used. This robust and independent collection was compiled in \cite{Nesseris:2017vor} and is complemented by \cite{Howlett:2017asq,Wang:2017wia,Zhao:2018gvb}. Measurements of WiggleZ \cite{Blake:2012pj}, SDSS-IV \cite{Zhao:2018gvb} and BOSS DR12 \cite{Wang:2017wia} each have a covariance matrix which we take into account, following \cite{Nesseris:2017vor,Sagredo:2018ahx}. }
\label{tab:rsd_data}
\end{table}

\begin{figure}[!]
\centering
\includegraphics[clip, trim = 0cm 17cm 0cm 0.9cm, width=0.9\linewidth]{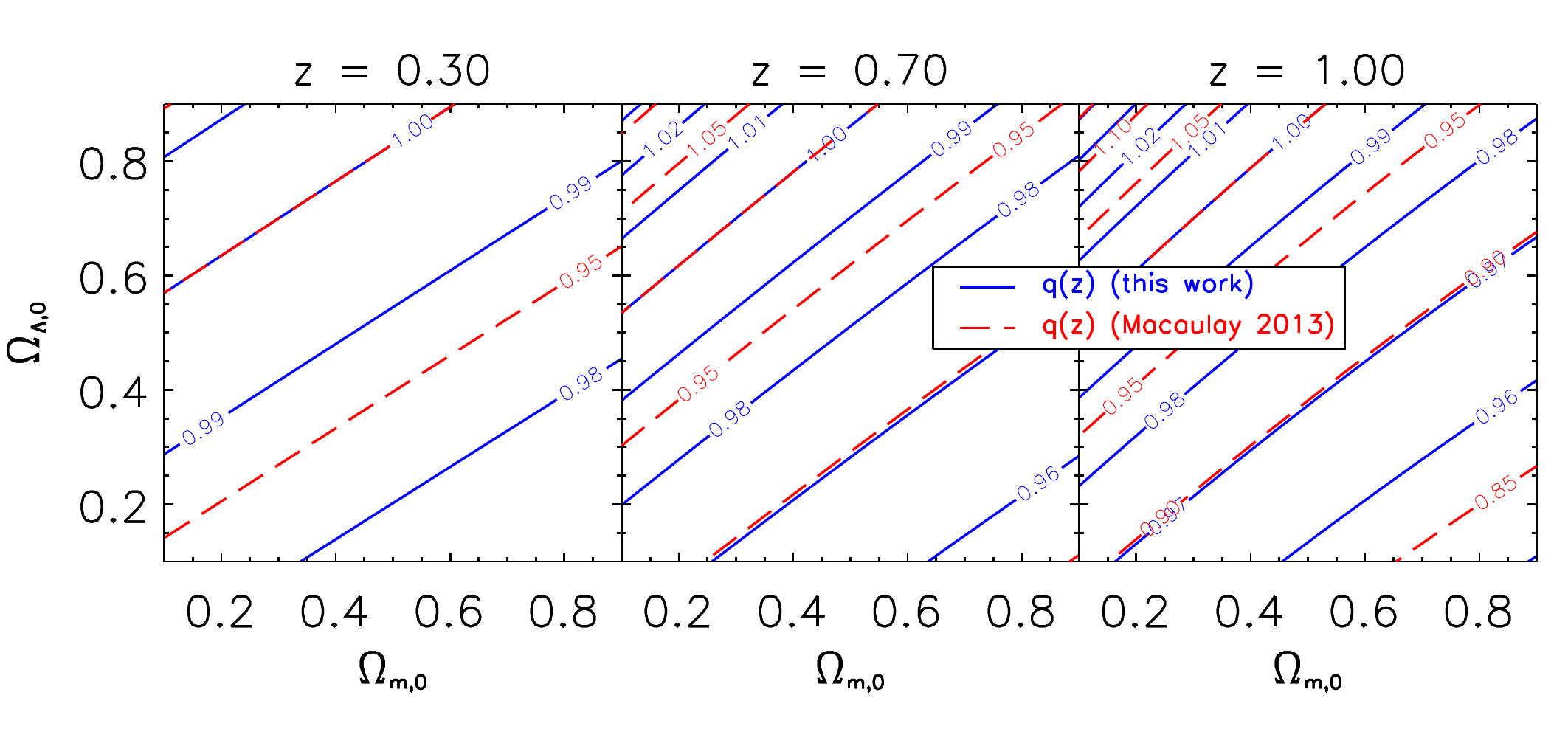}
\caption{AP rescaling factor $q(z)$ for $f\sigma_8$ at three redshifts as a function of the fiducial cosmology ($\Omega_{m,0}$, $\Omega_{\Lambda,0}$). Solid blue lines are our model \eqref{qus} (steps of 0.01); red dashed lines are \eqref{qth} (steps of 0.05) from \cite{Macaulay:2013swa}.} \label{fig:fs8ap}
\end{figure}

In \autoref{fig:fs8ap} we compare the AP rescaling factor $q$ of \eqref{qth} and \eqref{qus}, at three different redshifts. Although the degeneracy lines are identical, the amplitudes of the two corrections are significantly different, with the discrepancy increasing as a function of redshift. Note that our correction scheme predicts a smaller impact of geometric biases than the model in \cite{Macaulay:2013swa}. This in turn has non-negligible consequences for the statistical inference and cosmological parameter estimation processes.

~\\\textbullet~ {\bf Clustering Ratio}\\
Measurements of $\eta_R$ in \cite{Zennaro:2017qnp} are from BOSS DR7 \cite{SDSS:2008tqn} and BOSS DR12 \cite{SDSS-III:2015hof}, given in \autoref{tab:cr_data} and denoted CR in figures. These measurements used a flat $\Lambda$CDM fiducial model with $H_0 = 67$ km/s/Mpc, $\omo = 0.32$, a smoothing scale $R = 22h^{-1}$ Mpc and a correlation length $r = 2.1 R$. These were fixed using numerical simulations to optimise the signal-to-noise ratio. The advantage is that on these large scales the amplitude of the Clustering Ratio can be predicted using only linear perturbation theory. 

\begin{table}[!]
\small
\centering
\begin{tabular}{l c c c c}
\hline
Data set & $z_{\rm min}$ & $z_{\rm max}$ & $\eta_R$ & Ref. \\
\hline\hline
BOSS DR7 & 0.15 & 0.43 & 0.096 $\pm$ 0.007 & \cite{Zennaro:2017qnp,SDSS:2008tqn}\\
BOSS DR12 & 0.30 & 0.53 & 0.094 $\pm$ 0.006 & \cite{Zennaro:2017qnp,SDSS-III:2015hof}\\
BOSS DR12 & 0.53 & 0.67 & 0.105 $\pm$ 0.011 & \cite{Zennaro:2017qnp,SDSS-III:2015hof}\\ 
\hline
\end{tabular} 
\caption{Clustering Ratio data set used.}
\label{tab:cr_data}
\end{table}

As in the case of the RSD parameter, we map the `correct' $\eta_R$ computed in a given cosmology to $\tilde\eta_{\tilde R}$, which would be inferred in a `wrong' fiducial model. In \autoref{apcr} we show that the flat-space mapping \cite{Zennaro:2017qnp} applies also to curved-space models:
 \begin{equation}
\tilde\eta_{\tilde R} = \eta_{R}\qquad\mbox{where}\quad R = \alpha \tilde R\,.
\label{etatilde}
\end{equation}
Here $\alpha$ is the standard AP volume re-scaling factor: 
\begin{equation}\label{eq:def_alpha}
\alpha = \big( \alpha_\parallel\alpha_\perp^2 \big)^{1/3}. 
\end{equation}

\noindent\textbullet~ {\bf Hubble diagram of supernovae type Ia}\\ We use the Joint Light-curve Analysis sample compiled by \cite{SDSS:2014iwm}, denoted as SNIa in figures, which spans $ 0.01\leq z\leq 1.3$ and contains 740 SNIa. 
\\

\noindent\textbullet~ {\bf Baryon acoustic oscillations}\\ We use the BAO compilation of \cite{Planck:2018vyg} {(denoted `BAO' in figures)}, with measurements from 6dF \cite{Beutler:2011hx}, SDSS DR7 \cite{Ross:2014qpa} and BOSS DR12 \cite{BOSS:2016wmc}. 

The sound horizon at the end of the baryon drag epoch $r_{\rm d}$ is not fixed by CMB or priors. We treat the BAO scale as an uncalibrated standard ruler (whose amplitude depends on the physics of the early universe, and thus on the density parameters of baryons, CDM and neutrinos) to be adjusted using CMB-independent data. 
\\

\noindent\textbullet~ {\bf Cosmic microwave background}\\ We use the Planck 2018 low- and high-multipole temperature and polarisation power spectra \cite{Planck:2019nip}, denoting this data set as CMB in figures. In the analysis of our benchmark model, which includes spatial curvature, we set the relative amplitude of the lensing power parameter $A_{\rm L}$ to unity.

\subsection{Parameters and priors}\label{sec:priors}

\begin{table}
\small
\centering
\begin{tabular}{c c}
\hline\\[-1.5mm]
Parameter & Prior \\[1.5mm]
\hline\hline\\[-1.5mm]
$\omb h^2$              & $[0,\,100]$ \\[1mm]
$\omc h^2$              & $[0,\,100]$ \\[1mm]
$H_0$                   & $[40,\,100]$ \\[1mm]
$\tau$                  & $[0,\,0.2]$ \\[1mm]
$\ln(10^{10}A_{\rm s})$ & $[0,\,100]$ \\[1mm]
$n_{\rm s}$             & $[0.9,\,1]$ \\[1mm]
$\omk$                  & $[-0.2,\,0.6]$ \\[2mm]
\hline
\hline\\[-1.5mm]
$\omb h^2$              & $\mathcal{N}\left(0.0222,0.0005^2\right)$ \\[1mm]
$\sigma_{8,0}$          & $[0.6,\,1]$ \\[1mm]
$\omo$                  & $[0,\,1]$ \\[1.5mm]
\hline
\end{tabular} 
\caption{{\em Top:} $K\Lambda$CDM parameters and their corresponding prior range. {\em Bottom:} Additional priors considered in parts of our analysis.}
\label{tab:uniform_priors}
\end{table}

The benchmark for our analysis is the $K\Lambda$CDM model that we explore using a Monte Carlo Markov chain (MCMC) strategy by varying seven cosmological parameters: 
\begin{equation}\label{eq:free_params}
\Big\lbrace \omb h^2,\; \omc h^2,\; H_0,\; \tau,\; \ln(10^{10}A_{\rm s}),\; n_{\rm s},\; \omk \Big\rbrace.
\end{equation}

We use the Boltzmann code\footnote{\url{www.class-code.net}} \texttt{CLASS} \cite{Blas:2011rf} for cosmology and the ECLAIR suite \cite{Ilic:2020onu} for sampling likelihoods. 

We present results on the growth of structure using $\sigma_{8,0}$ and the combination $S_8=\sigma_{8,0}(\omo/0.3)^{0.5}$ (e.g. \cite{Kilbinger:2012qz, KiDS:2020suj}) that better resolves the degeneracy between $\sigma_{8,0}$ and $\omo$. The constraints from our analysis are model-dependent and need to be understood in combination with our priors and assumptions. We assume the minimal neutrino scenario (one with mass 0.06\,eV, two massless). We try to be as agnostic as possible when imposing conditions on variation of free parameters and impose the very large uniform priors displayed in \autoref{tab:uniform_priors}. Consequently, the full cosmological potential of each probe can be better assessed. 

We assume an informative Gaussian prior on the baryon density, $\omb h^2 = 0.0222 \pm 0.0005$, as set by primordial deuterium abundance measurements \cite{Cooke:2017cwo} {(denoted `BBN' in figures)}. This choice ensures that our conclusions are completely independent of CMB measurements. However, we note that estimation of $\omb h^2$ from the shape of CMB acoustic peaks is virtually independent of the model, i.e. essentially insensitive to the late-time expansion and early-time physics \cite{Audren:2012wb}. 

\section{Cosmological constraints on $K\Lambda$CDM}\label{sec4}

Planck observations \cite{Planck:2018vyg} have resulted in some unexpected results with important phenomenological consequences for the standard flat-space six-parameter $\Lambda$CDM model. It appears that if $\Lambda$CDM is extended to include spatial curvature, then the Planck temperature and polarisation maps favour (statistically significant) negative values (i.e. $K>0$): $\omk = -0.044\,^{+0.018}_{-0.015}$. This does not leave the rest of the model unchanged. Indeed, it implies $H_0 = 54.4\,^{+3.3}_{-4.0}$ at $68\%$ CL, much lower than direct measurements by virtually all cosmological probes. In addition, it suggests a matter content $\omo$ about $50\%$ larger than expected and generally measured \cite{DiValentino:2019qzk}. 

A driver for the $K>0$ result is that the Planck temperature power spectrum is best fitted by models with significantly more lensing distortions than actually observed. The nonphysical lensing factor, if left free, is constrained to values $A_{\rm L}>1$ \cite{Planck:2013pxb} in the flat $\Lambda$CDM model. A more physical parameter that can affect the lensing is in fact the curvature density: more negative $\omk$ leads to larger $A_{\rm L}$ \cite{DiValentino:2019qzk}. 
The central goal of our analysis is the characterisation of the curvature in one-parameter extensions of the flat $\Lambda$CDM scenario. We therefore take $K\Lambda$CDM as the reference model and show how, by jointly exploiting different techniques to extract clustering information (Clustering Ratio and RSD), we can gain insights into this issue.

\subsection{Constraints from low-redshift clustering probes}\label{sec4.1}

\begin{figure}[!]
\begin{center}
\includegraphics[width=0.8\linewidth]{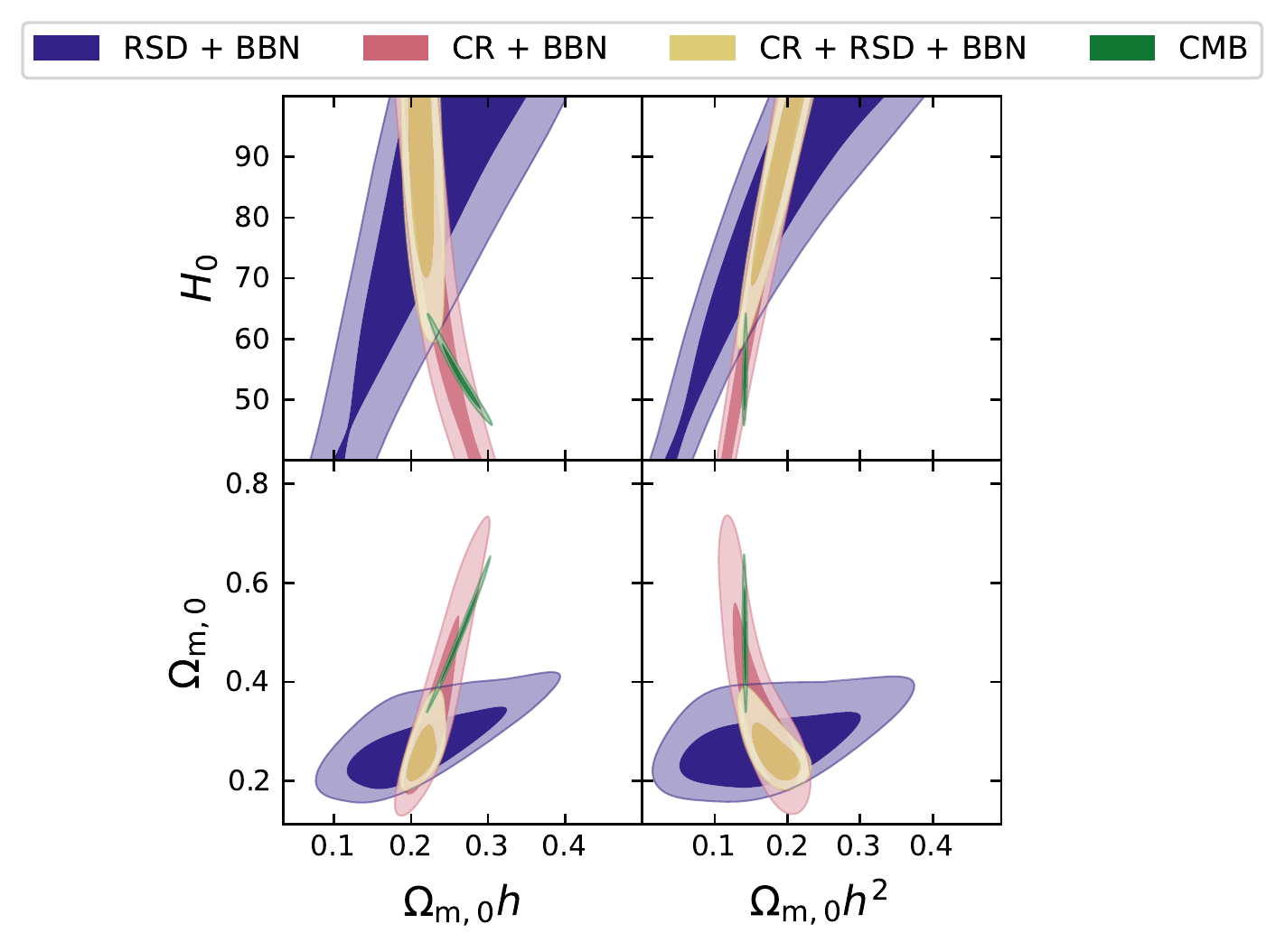} 
\caption{Marginalised 2D posterior distributions of some relevant parameters of $K\Lambda$CDM, from combinations of RSD compilation + BBN prior (purple), CR compilation + BBN prior (pink), CR + RSD compilations + BBN prior (yellow), CMB temperature + polarisation (green). Contours are 68\% and 95\% CL.}
\label{fig:clust}
\end{center}
\end{figure}

Constraints on some parameters of $K\Lambda$CDM derived from low-redshift ($z\lesssim 2$) CR and RSD clustering observables are shown in \autoref{fig:clust}. The measurements are performed on non-overlapping data\footnote{For example, the CR and RSD data sets both contain BOSS DR12 measurements, which we consider only in the CR data set when making any CR + RSD combinations, to avoid double-counting.}, and moreover probe the universe on very different epochs and scales. The resulting likelihood contours are thus statistically independent and can be meaningfully compared. 

Regarding the constraining power of individual probes we see that the Clustering Ratio performs extremely well in constraining the combination $\omo h$:
\begin{equation}
 \omo h=0.232\,_{-0.040}^{+0.057} \qquad (K\Lambda\text{CDM}: \;\;\; \text{CR}),
\end{equation}
at 95\% CL. This is comparable to the constraining power of CMB alone ($\omo h = 0.264\,_{-0.035}^{+0.033}$) and improves by a factor of three on what can be achieved with RSD analyses ($\omo h=0.214\,_{-0.106}^{+0.141}$). This power results from the degeneracy axis in the $(\omo, h)$ plane along the line $\omo h =\,$const, a fact that was already noticed and discussed by \cite{Bel:2013csa} and \cite{Bel:2013ksa}. The preferred degeneracy line is physically due to the fact that, for large values of $R$, the CR parameter $\eta_R$ is only sensitive to the shape of the matter power spectrum, which itself depends in a degenerate way on the product $\omo h$. 

As shown by \cite{Bel:2013ksa}, the degeneracy axis in the $(\omo,\, h)$ plane develops along the line $\omo h^2 = \,$const if the CR analysis is performed in the mildly nonlinear regime, i.e. for small smoothing length $R$. Thus in principle, additional constraining power from clustering probes alone could be achieved by combining RSD and CR (for a given fixed scale $R$) on a larger redshift interval than currently available. Future DESI and Euclid data could make this possible.

By contrast, RSD are more effective in constraining $\omo$ as evident in the lower panels of \autoref{fig:clust} -- since RSD are sensitive to normalisation of the power spectrum, $\sigma_{8,0}$. The CR, at least when measured on linear scales (i.e. by smoothing the galaxy field using large values of $R$), is insensitive to the power spectrum amplitude. Denser samples would be needed to perform a small-scale analysis (in slightly nonlinear regimes) capable of extracting information about $\omo$ also from CR measurements alone. 

We also note the different statistical degeneracies displayed in parameter space by the clustering probes, with the misalignment of the degeneracy axes forcing the likelihood contours to overlap in a narrow region. Although the cosmological resolution provided by the clustering probes alone is weaker than the CMB, their combination is effective in placing some CMB-independent constraints on $H_0$ and $\omo$.

The parameter $\omo h$ is constrained by the joint clustering analysis as $\omo h = 0.216 \pm 0.012$ ($68\%$ CL). This figure improves on the constraint $\omo h =0.220\,^{+0.029}_{-0.019}\,$ that is obtained by \cite{Brieden:2021cfg}, using a similar approach (i.e. including the shape of the power spectrum in the analysis); however, we note that \cite{Brieden:2021cfg} includes information about the BAO, makes a zero-curvature assumption and leaves $\Omega_{\rm b,0} h^2$ as a free fitting parameter. When compared to the results of \cite{Ivanov:2019pdj}, who found that the combination $\omo h^2$ can be measured directly from the full-shape analysis of the BOSS power spectrum (plus the BBN and the flatness constraint $K=0$) with $5\%$ precision (at 68\% CL), these results strengthen the evidence that clustering at low redshifts can be a source of independent measurements whose accuracy rivals that of the Planck CMB data.

The linear perturbation analysis of clustering at $z \lesssim 2$ places a lower bound $H_0>62$ km/s/Mpc ($95\%$ CL). Although far from resolving the current indeterminacy in the precise value of $H_0$, this constraint has interesting conceptual and practical implications. We highlight that it provides a fully background-independent limit on $H_0$ -- which is bounded from below without exploiting geometric probes such as distances, areas or volumes, nor calibration ladders, but by relying only on the linear measurements of clustering (CR and RSD) and on the primordial abundance of baryons set by BBN measurements (the prior on $\omb h^2$). 

In the presence of tensions within the standard cosmological paradigm, it is reassuring that this `excluded zone' in parameter space is consistent with the results from cosmological probes of the background. This result also allows conclusions to be drawn about the magnitude of other fundamental parameters of the model, for example by imposing limits on the possible curvature of the spatial sections of the universe (see next section). Indeed, it assists with resolving the geometric degeneracy of the CMB \cite{Bond:1997wr, Zaldarriaga:1997ch}, i.e. the fact that changes in $\omk$ can be compensated by changes of $H_0$ and produce thereby almost no observational effects on the angular diameter distances to the last scattering surface \cite{Planck:2015fie}. 

Similarly, interesting constraints on the matter density emerge from our analysis. Without fixing the spatial curvature and considering only clustering information (supplemented by the BBN prior on $\omb h^2$), we find that {at $68\%$ CL}
\begin{equation}
 \omo=0.26\pm 0.04 \qquad (K\Lambda\text{CDM}: \;\;\; \text{CR + RSD + BBN}).
\end{equation}
At $95\%$ CL, we obtain $0.203 \leq \omo \leq 0.370$. We have run additional MCMC to assess the range of validity of this result. We verify that the upper bound of $\omo$ is not forced by the lower bound of $\sigma_{8,0}$. To this end we reanalyse the data by widening the interval of the flat prior on $\sigma_{8,0}$, finding that this does not induce any appreciable change in the credible interval of $\omo$. 

Next, we relax the BBN constraint by setting a flat prior on $\omb h^2$ in the interval $[0,100]$. Contrary to what happens for the lower bound on $H_0$ (which is degraded), the constraint on $\omo $ is virtually unchanged. This is {\em indisputable evidence of the presence of dark matter from only low-z clustering data} -- independent of any primordial constraints on power spectrum amplitude, normalisation and shape.

We also find that statistically insignificant changes are introduced in the $H_0$ and $\omo$ constraints when the neutrino mass is set to zero. These constraints are confirmed both qualitatively and quantitatively when the joint clustering analysis is performed by adding also the RSD measurements from the DR7+DR12 samples, i.e. the data sets for which measures of the CR are available and the sample independence condition fails. 

The analysis of galaxy clustering alone is not effective in placing useful constraints on the curvature parameter (and also on the primordial scalar spectral index $n_{\rm s}$). For the CR, the reason is the limited range of spatial scales surveyed by currently available galaxy samples. In fact, \autoref{fig:snu} shows that the shape of the matter power spectrum becomes sensitive to spatial curvature only scales comparable to the horizon. For $f\sigma_8$, the reason is that the scale-independent curvature effects induced by the specific evolution of the linear growth rate (see \autoref{fig:even}) are degenerate with the amplitude (normalisation) $A_{\rm s}$ of the power spectrum and thus cannot be constrained. 

\subsection{Combining clustering with geometric probes (BAO and SNIa)}\label{sec4.2}

\begin{table}[!]
\small
\centering
\begin{tabular}{c r c}
\multicolumn{3}{c}{$K\Lambda$CDM model}\\[1mm]
\hline \\[-2.3mm]
 Comparison & $ \log_{10}I $ & Agreement / Disagreement\\[1.5mm]
\hline\hline\\[-2mm]
CR + BBN vs RSD + BBN & 1.35  & strong agreement \\[1mm]
SNIa vs RSD + BBN     & 0.48   & inconclusive \\[1mm]
SNIa vs CR + BBN      & 0.14   & inconclusive \\[1mm]
BAO vs CR + BBN       & 0.07   & inconclusive \\[1mm]
BAO vs RSD + BBN      & 0.04   & inconclusive \\[1mm]
CMB vs CR + BBN       & $-0.28$ & inconclusive \\[1mm]
CMB vs SNIa           & $-0.7$  & substantial \\[1mm]
CMB vs BAO            & $-2.14$ & decisive disagreement \\[1mm]
CMB vs RSD + BBN      & $-2.75$ & decisive disagreement  \\[1mm]
\hline
\end{tabular} 
\caption{Consistency between data sets in $K\Lambda$CDM (first column), the corresponding $ \log_{10}I $ (middle column) and interpretation on Jeffrey's scale (right column).} 
\label{tab:agreement}
\end{table}

\begin{figure}[!]
\begin{center}
\includegraphics[width=0.9\linewidth]{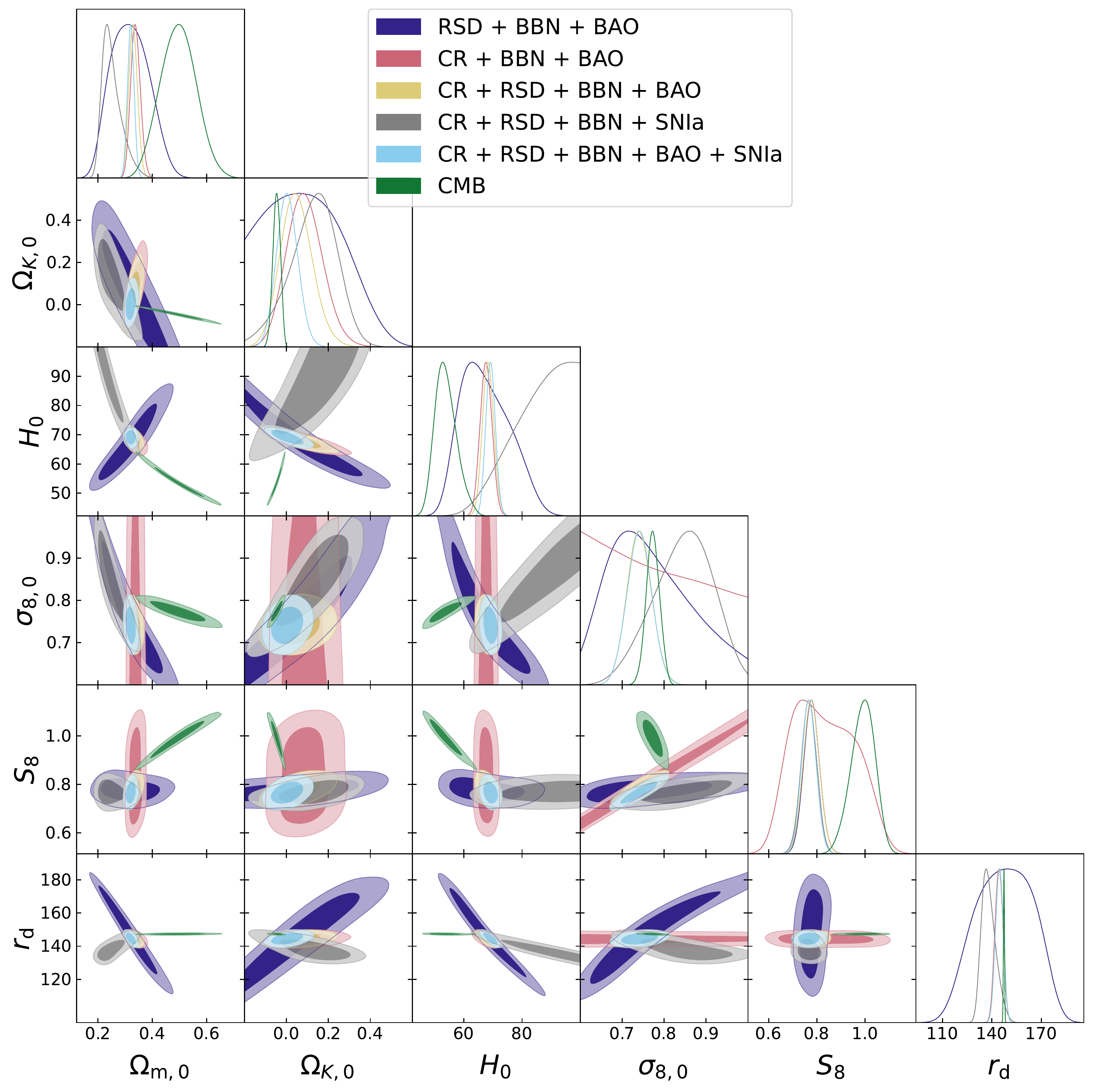}
\caption{As in \autoref{fig:clust}, with different data sets and $K\Lambda$CDM parameters.}
\label{fig:clust_bg}
\end{center}
\end{figure}

It is necessary to combine clustering with independent probes if we are to increase the predictive power on spatial curvature. First we add geometric information from the location of the BAO peak, a background observable which can be extracted from spectroscopic surveys. This provides the ideal test bed to fully evaluate the cosmological potential of galaxy redshift surveys. Then we include the Hubble diagram of SNIa. In this case, the goal is slightly different: to evaluate the efficiency of CMB-independent, low-redshift data in placing competitive constraints on cosmological parameters. 

A critical requirement is consistency of the different data sets, i.e. they need to be compatible with being drawn from the same set of cosmological parameters. Only then can they be jointly analysed in a meaningful way to increase the statistical power of the inference. It is therefore essential to investigate and verify whether this condition is realised in practice. Several statistical methods have been proposed to quantify the level of concordance (or discordance) between two independent cosmological data sets $D_1$ and $D_2$ (e.g. \cite{Marshall:2004zd, Raveri:2015maa, Joudaki:2016kym}). 

We adopt the approach of \cite{Joudaki:2016kym}, based on the Deviance Information Criterion (DIC) \cite{Spiegelhalter:2002yvw}, calculated as $2\overline{\chi_{\rm eff}^2}-\chi^2_{\rm eff}$, where $\chi^2_{\rm eff} =-2 \ln \mathcal{L}_{\rm max}$ and $\mathcal{L}_{\rm max}$ is the maximum likelihood of the data given the model. The overbar denotes the mean taken over the posterior distribution.
Then we evaluate the statistic 
\begin{eqnarray}
 {I}(D_1,D_2)= {\rm e}^{-\mathcal{F}(D_1,D_2)/2}~~\mbox{where}~
\mathcal{F}(D_1,D_2)={\rm DIC}(D_1 \cup D_2)-{\rm DIC}(D_1)-{\rm DIC}(D_2).
\end{eqnarray}
This is constructed so that there is concordance between the data sets when $\log_{10} I$ is positive, and discordance when it is negative. More precisely, according to the Jeffrey scale, the agreement/ disagrement is considered `substantial' if $|\log_{10} I | >0.5, $ `strong' if $|\log_{10}I|>1.0$ and `decisive' if $|\log_{10}|>2.0$. 
 
\autoref{tab:agreement} shows the degree of concordance/ discordance in the various sample pairs considered in our analysis. There is strong agreement between the CR and RSD samples, which justifies the joint analysis presented in the previous section. Also, clustering and background data sets are not incompatible. This suggests that {\em CR, RSD, BAO and SNIa can be combined in a meaningful way even when the curvature parameter is free to vary} in the likelihood analysis. \autoref{fig:clust_bg} shows the $68\%$ and $95\%$ CL constraints for various parameters. It demonstrates that the residual degeneracy affecting the likelihood contours derived from combining BAO + RSD is substantially resolved when the CR probe is added. The most dramatic improvement is in $H_0$ and $\omk$ (see \autoref{tab:baosnia}). Even higher precision is reached if we include SNIa data in the likelihood analysis. 

\begin{table}[!h]
\small
\centering
\begin{tabular}{l c c}
\multicolumn{3}{c}{$K\Lambda$CDM model}\\[1mm]
\hline \\[-2mm]
 & Clustering+BBN+BAO & Clustering+BBN+BAO+SNIa \\[1.5mm]
\hline\hline\\[-2mm]
$H_0$\,(km/s/Mpc)    & $68.32 \pm 1.92$               & $67.55\,_{-1.63}^{+1.60}$\\[2mm]
$\omo$               & $0.3299\,_{-0.0158}^{+0.0157}$ & $0.3210 \pm 0.0121$\\[2mm]
$\omk$               & $0.0484\,_{-0.0756}^{+0.0748}$ & $0.0041\,_{-0.0504}^{+0.0500}$\\[2mm]
$n_{\rm s}$          & $0.950 \pm 0.034$              & $0.951 \pm 0.034$\\[2mm]
$\sig80$             & $0.741 \pm 0.029$              & $0.741 \pm 0.029$\\[2mm]
$S_8$                & $0.777 \pm 0.033$              & $0.766 \pm 0.030$\\[2mm]
$r_{\rm d}$\,(Mpc)   & $144.50\,_{-2.35}^{+2.33}$     & $144.57 \pm 2.34$\\[2mm]
\hline
\end{tabular} 
\caption{Marginalised mean and 68\% CL on some parameters of the $K\Lambda$CDM model. Constraints are from the combination of clustering (CR + RSD) data (with BBN prior) and background probes: BAO (middle column) and BAO+SNIa (right column).} 
\label{tab:baosnia}
\end{table}

The most interesting result is that {\em there is convincing evidence, independent of CMB data, for not rejecting the hypothesis that the Universe is spatially flat}. The mean curvature parameter derived using Clustering + BBN + BAO + SNIa data is 
\begin{equation}
\omk=0.0041\,_{-0.0504}^{+0.0500} \qquad (K\Lambda \text{CDM:} \;\;\; \text{CR + RSD + BBN + BAO + SNIa}), 
\end{equation}
at $68\%$ CL. Note that this evidence in favour of a flat-space Universe is stable and does not change if SNIa data are not included in the likelihood analysis: in this last case we obtain $\omk=0.0484\,_{-0.0756}^{+0.0748}$. Our finding confirms and strengthen the results of \cite{Chudaykin:2020ghx}, which uses similar external information (BBN prior, SNIa+BAO data), but a different clustering probe (full-shape analysis of the BOSS power spectrum) to obtain $\omk = -0.043\,^{+0.036}_{-0.036}$ (68\% CL). 

We also obtain, at $68\%$ CL, 
\begin{eqnarray}
&& \omo = 0.3210 \pm 0.0121\,, \quad
H_0 = 67.55\,_{-1.63}^{+1.60}~
\text{km/s/Mpc}
\notag\\
&& (K\Lambda \text{CDM: ~CR + RSD + BBN + BAO +SNIa}),
\end{eqnarray}
which are fully compatible with constraints obtained by  fitting the flat $\Lambda$CDM six-parameter baseline model to  clustering+BAO+BBN data  \cite{Brieden:2021cfg} ($H_0= 66.0\,^{+2.0}_{-1.7}$~km/s/Mpc) -- despite the fact that curvature is treated in our analysis as a free fitting parameter. Our results are also compatible with fitting the concordance model to Planck CMB power spectra (TT, TE, EE$+$lowE): $\omo = 0.3166 \pm 0.0084$ and $H_0=67.27 \pm 0.60$~km/s/Mpc. We find that {\em the mean value of $H_0$, whether or not the SNIa sample is used, is lower than the local measurement of $H_0$}, using the distance-ladder technique. Given the lower resolution (roughly three time worse) than the CMB estimate, the tension with the latest measurement, $H_0=73.04\pm 1.04$~km/s/Mpc, from the Cepheid-SN sample of \cite{Riess:2021jrx}, is reduced to $\sim 2.85 \sigma$ from the $\sim 4.8\sigma$ discrepancy with the Planck best fitting value. 

By adding the BAO prior we create a connection with the physics of the early Universe, via the sound horizon. However, we stress that we do not fix this scale using CMB information, but rather leave it as a free fitting parameter to be determined by low-redshift data. As a consequence, an interesting by-product of our analysis is {\em a CMB-independent measurement of the acoustic horizon at the drag epoch, obtained without assuming spatial flatness:} 
\begin{equation}
 r_{\rm d} = 144.50\,_{-2.35}^{+2.33} ~~ {\rm Mpc} \qquad (K\Lambda\text{CDM:} \;\;\; \text{ CR + RSD + BBN + BAO}),
\end{equation}
at 68\% CL. This estimate is in fair agreement with the value from a similar analysis by \cite{Chudaykin:2020ghx}: $r_{\rm d}=146\pm 2.4$~Mpc. Note that the $r_{\rm d}$ value strongly depends on the BBN prior on $\Omega_b h^2$; indeed, with a loose non-informative flatness prior, we find $r_{\rm d} = 142.85\,_{-15.74}^{+17.85}$~Mpc. 

\subsection{Combining clustering with the CMB}\label{sec4.3}

Combining low-redshift data sets with the Planck likelihood in $K\Lambda$CDM is a more critical step, as most samples are incompatible with the CMB data. This issue was highlighted by \cite{DiValentino:2019qzk}, which showed that, although the joint analysis of various probes with CMB favours a flat-space model, the results should be viewed with suspicion as they are obtained by combining data sets whose individual constraints are not in principle compatible.

Before the combination issue is tackled, a preliminary assessment is needed to test whether the Planck measurement $\omk = -0.044\,^{+0.018}_{-0.015}$ rules out standard $\Lambda$CDM. At first glance, it is clear that CL intervals on individual variables (obtained after marginalising over all the remaining parameters), or projected bi-dimensional contours, seem to exclude the standard $\Lambda$CDM. However, the integration of unnecessary degrees of freedom into a multi-dimensional parameter space involves projection and volume selection effects. For example, a model might fall outside a given CL defined by the posterior after the inclusion/ exclusion region is determined in a lower-dimensional subspace of the full parameter space $S$ -- although the same model may be inside the very same CL once the credible region is calculated in the full volume $S$. 

This is indeed what happens for the Planck data. The best fitting flat $\Lambda$CDM model of Planck appears to be outside the $95\%$ CL, if judged by looking only at the CL of $\omk$, or the 2D CLs on various pairs of parameters shown in \autoref{fig:clust_bg}. However, the best fitting $\omk=0$ Planck model (which has a $\chi^2 =2773.55$) is still substantially within the 68\% CL of the full $K\Lambda$CDM parameter space, which is identified by the iso-contour $\chi^2=2785.5$. Nevertheless, one should not ignore the fact that if the best fitting $K\Lambda$CDM and flat $\Lambda$CDM models are compared using Bayesian hypothesis testing, the curved model, despite its additional degree of freedom, is strongly favoured by Planck data \cite{DiValentino:2019qzk}.

It is therefore worth assessing whether the latter result remains unchanged after combining CMB with CR data. Indeed, as is evident in \autoref{tab:agreement}, CR is the only low-$z$ sample that, according to Bayesian evidence as measured by DIC, does not disagree with the CMB data given the $K\Lambda$CDM model. This compatibility becomes even more important when compared to the fact that most of the low-$z$ samples, both BAO and full-shape power spectrum measurements, are incompatible with the CMB data in the $K\Lambda$CDM framework and, therefore, the constraints obtained from their combination must be viewed with caution \cite{Vagnozzi:2020rcz}. 

\begin{figure}[!]
\begin{center}
\includegraphics[width=0.9\linewidth]{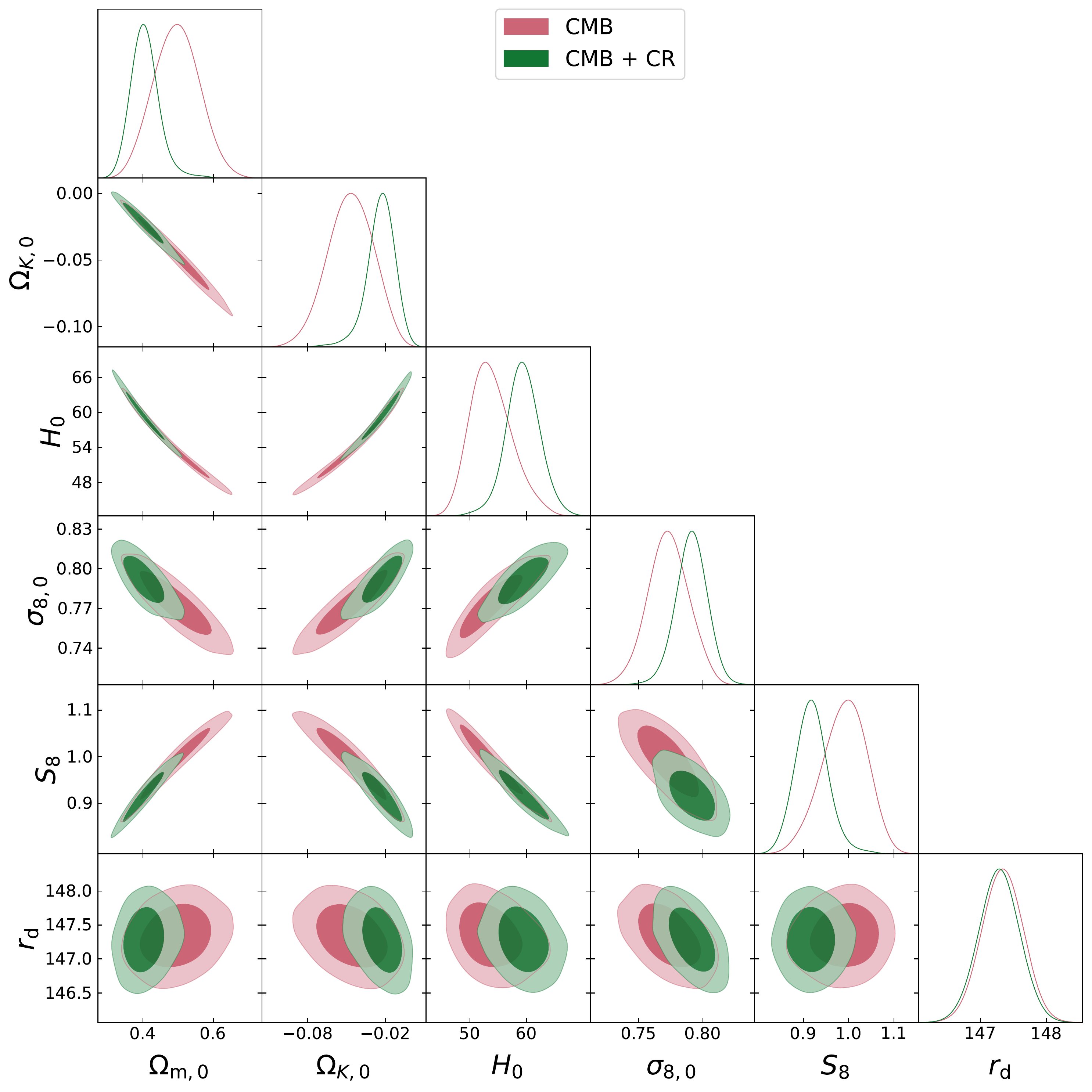}
\caption{As in \autoref{fig:clust_bg}, but combining CR only with CMB.}
\label{fig:clust_cmb}
\end{center}
\end{figure}
 
Results from the joint analysis of CMB + CR data sets compared to Planck alone are displayed in \autoref{fig:clust_cmb}. We do not show constraints on $n_{\rm s}$, $A_{\rm s}$ and $\omc$, since these parameters are essentially determined by Planck data alone, with marginal contribution from CR. On the contrary, we find that CR + CMB has significant impact on the estimate of curvature: 
\begin{equation}
\omk= -0.023 \pm 0.010 \qquad (K\Lambda\text{CDM:} \;\;\; \text{CR + CMB}),
\end{equation}
at 68\% CL. This measurement brings the flat $\Lambda$CDM model back into agreement within $\sim 2\sigma$.

To assess the preference for either model, instead of directly comparing their $\chi^2$, we use again the DIC, as suggested in \cite{Kunz:2006mc, Joudaki:2016kym}, an indicator that takes Bayesian complexity into account and allows comparison of models with a different number of parameters. As reference, a difference $\Delta$DIC\,$= 10(5)$ is conventionally interpreted as strong (moderate) preference for the model with lower DIC. If the difference is even smaller, the statistic is inconclusive. As in the classical likelihood ratio test, in comparing an extended model (here $K\Lambda$CDM) with a reference model ($\Lambda$CDM), we take positive values of $\Delta$DIC to indicate a preference for the extended model.

When CMB data alone are used to compare $K\Lambda$CDM and $\Lambda$CDM, \cite{DiValentino:2019qzk} find $\Delta$DIC\,$\sim 7.4$, which is degraded from $\Delta \chi^2_{\rm eff} \sim 11$ due to the increased Bayesian complexity. This is interpreted as $K\Lambda$CDM preferred over $\Lambda$CDM with an odds ratio of about 41\,:\,1. If instead the CMB+CR data are used to compare the two models, we find $\Delta$DIC\,$\sim 4.5$ (degraded from $\Delta \chi^2_{\rm eff} \sim 9$), i.e. inconclusive evidence in favour of $K\Lambda$CDM. Put differently, in terms of betting probability, $K\Lambda$CDM is preferred to the $\Lambda$CDM with odds of 10\,:\,1, a value too low to reject flatness without too much risk of being wrong.

Similar to what happens for $\omk$, other parameters also deviate significantly from the Planck best-fit value once CMB and CR data are jointly analysed. For example, CMB + CR gives $H_0 = 59.33\,_{-2.81}^{+2.93}$~km/s/Mpc at 68\% CL. 

\section{Constraints on the flat $\Lambda$CDM model}\label{sec5}

\begin{figure}[h]
\begin{center}
\includegraphics[width=0.8\linewidth]{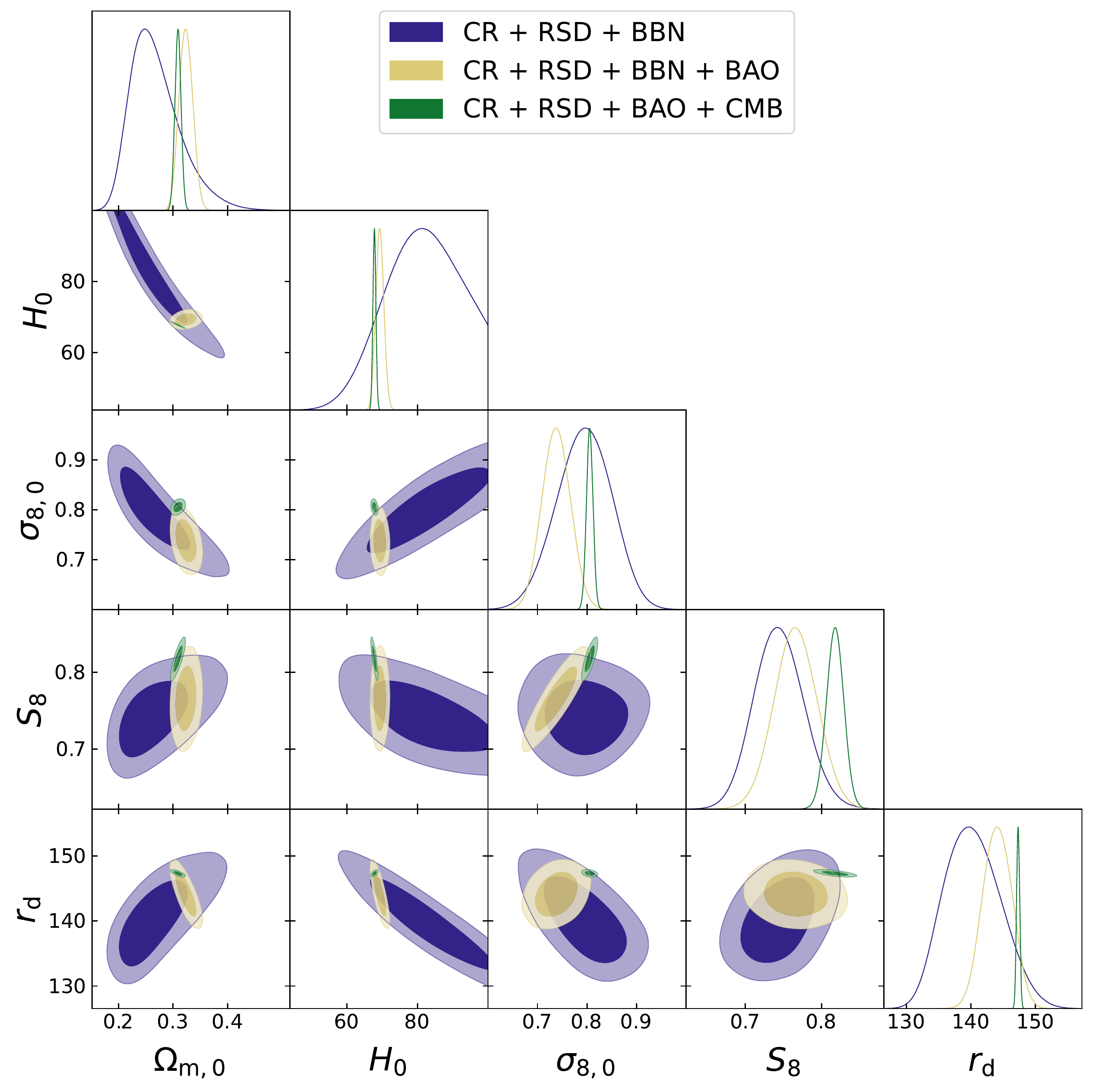}
\caption{As in \autoref{fig:clust_bg}, but for $\Lambda$CDM. }
\label{fig:flatlcdm}
\end{center}
\end{figure}

\begin{table}
\small
\centering
\begin{tabular}{l c c c}
\multicolumn{4}{c}{Flat-$\Lambda$CDM model}\\[1mm]
\hline \\[-2mm]
 & Clustering + BBN & Clustering + BBN + BAO & Clustering + BAO + CMB \\[1.5mm]
\hline\hline\\[-2mm]
$H_0$\,(km/s/Mpc)  & $81.22\,_{-10.61}^{+11.17}$    & $69.31\,_{-1.15}^{+1.14}$             & $67.80 \pm 0.42$  \\[2mm]
$\omo$             & $0.2680 \pm 0.0451$            & $0.3236\,_{-0.0126}^{+0.0128}$          & $0.3090 \pm 0.0056$ \\[2mm]
$\omo h^2$         & $0.1726\,_{-0.0196}^{+0.0198}$ & $0.1556\,_{-0.0089}^{+0.0088}$ & $0.1420 \pm 0.0009$ \\[2mm]
$n_{\rm s}$              & $0.949 \pm 0.034$             & $0.947\,_{-0.033}^{+0.034}$     & $0.967 \pm 0.004$\\[2mm]
$\sigma_{8,0}$     & $0.794 \pm 0.056$             & $0.734 \pm 0.029$            & $0.806 \pm 0.007$ \\[2mm]
$S_8$              & $0.744 \pm 0.032 $           & $0.766 \pm 0.028$              & $0.818 \pm 0.011$ \\[2mm]
$r_{\rm d}$\,(Mpc) & $140.31\,_{-4.47}^{+4.44}$   & $144.11\,_{-2.27}^{+2.29}$     & $147.31 \pm 0.23$ \\[2mm]
\hline
\end{tabular} 
\caption{As in \autoref{tab:baosnia}, but for $\Lambda$CDM -- and therefore including a CMB combination (last column).}
\label{tab:flat}
\end{table}

The central values of $K\Lambda$CDM parameters in \autoref{tab:baosnia} are all consistent with the Planck best fitting $\Lambda$CDM model at 68\% CL (see Table 2 of \cite{Planck:2018vyg}). We interpret this as clear evidence that {\em the current low-redshift CMB-independent data do not prefer {curved} extensions over the standard flat $\Lambda$CDM model}. 

There has been wide-ranging discussion, within the framework of $\Lambda$CDM, about the tendency of late-time measurements to prefer slightly lower matter density or clustering amplitude compared to early-Universe CMB measurements (see e.g. \cite{HSC:2018mrq, DES:2017myr, DES:2018rvo, KiDS:2020suj} and \cite{Perivolaropoulos:2021jda} for a review). As new clustering probes becomes available, like the Clustering Ratio, we can explore further this issue and determine whether there are indications of convergence towards or away from the Planck CMB prediction.

Our analysis, summarised in \autoref{fig:flatlcdm} and \autoref{tab:flat}, confirms that perturbation probes such as RSD and CR (together with the BBN prior) favour a slightly lower value of the matter density: 
\begin{equation}
\omo = 0.2680 \pm 0.0451     \qquad (\Lambda\text{CDM:} \;\;\; \text{CR + RSD + BBN}).
\end{equation}
For geometric probes: 

$\omo=0.353\pm 0.037$ (SNIa \cite{FSS:2018cey}); 

$\omo=0.301 \pm 0.016$ (BAO \cite{eBOSS:2020yzd}); 

$\omo= 0.3166 \pm 0.0084$ (CMB \cite{Planck:2018vyg}).

\noindent Our estimate is in excellent agreement with various other clustering probes, for example: 

$\omo=0.295\pm 0.01$ from the BOSS galaxy power spectrum (with BBN prior) \cite{Ivanov:2019pdj}; 
$\omo=0.276 \pm 0.047$ from SPT galaxy clusters (SZ, WL, X-ray) \cite{SPT:2018njh}; 

$\omo={0.28}\,^{+0.05}_{-0.04}$ from galaxy clusters in KiDS DR3 \cite{Lesci:2022owx}. 

\noindent It is also consistent with: 

$\omo = 0.22\,^{+0.05}_{-0.04}$ from joint analysis of SDSS clusters and weak lensing \cite{DES:2018crd}; 

$\omo=0.290\,^{+0.039}_{-0.063}$ from DES-Y3 weak lensing \cite{Amon:2022ycy}; 

$ \omo= 0.264\,^{+0.032}_{-0.019}$ from 3$\times$2 analysis (shear, galaxy counts, galaxy-shear) on DES-Y1 \cite{DES:2017myr}; 

$\omo=0.339\,^{+0.032}_{-0.031}$ from 3$\times$2 analysis repeated on DES-Y3 \cite{DES:2021wwk}.

\noindent It is interesting to compare these values in $\Lambda$CDM, with the value we find by relaxing the flatness assumption. Contrary to what happens in CMB analysis, where $\omo$ is revised upwards when curvature is allowed, we find that $\omo$ is systematically lower than the `geometric' estimates also in $K\Lambda$CDM: $\omo= 0.26\,\pm 0.04$ (see \autoref{fig:clust}). We conclude that {\em the low $\omo$ values obtained from different kinds of clustering analysis are unlikely to arise from setting $\omk=0$.}

The growth parameter 
\begin{equation}
S_8 = 0.744 \pm 0.032  \qquad (\Lambda\text{CDM:} \;\;\; \text{CR + RSD + BBN})
\end{equation}
is about $2.5\sigma$ lower than measured by Planck ($S_8=0.834\pm 0.016$) but in agreement with other late-time probes, notably lensing studies

$S_8 = 0.759\,^{+0.025}_{-0.023}$ from shear in DES-Y3 \cite{Amon:2022ycy}; 

$S_8 = 0.759\,^{+0.024}_{-0.021}$ from shear in KIDs-1000 \cite{KiDS:2020suj}

$S_8 = 0.766\,_{-0.014}^{+0.020}$ from shear and RSD in KIDs-1000 \cite{Heymans:2020gsg} 

\noindent and clustering analysis of the BOSS DR12 dataset ($S_8 = 0.751 \pm 0.039$ 
\cite{Philcox:2021kcw}, $S_8=0.736 \pm 0.051$ \cite{Chen:2021wdi}).
\noindent This result is stable also when we enlarge the parameter space to include curvature (see \autoref{tab:baosnia}).

It is tempting to interpret the analysis as reinforcing the evidence for a possible tension that is not resolved by spatial curvature, and thus suggesting a scenario in which gravity is weaker than predicted by $\Lambda$CDM model. However, it is worth sticking to a strict interpretation of statistical significance. Data carry central values but also statistical fluctuations. \autoref{fig:clust_bg} and \autoref{fig:clust_cmb} show that nonzero curvature is more likely to be a fluke than a physical signal that indicates a tension between $\omo$ measurements in the early and late Universe. This conclusion is supported by the fact that CMB data are known to provide a sub-optimal estimate of spatial curvature, due to the `geometric degeneracy' between cosmological parameters \cite{Bond:1997wr, Efstathiou:1998xx}. Precise measurements at arc-minute scales made possible an estimate of spatial curvature from a single CMB experiment -- but this measurement is made at a redshift where the spatial geometry does not directly influence the physics of matter clustering. Similarly to the case of curvature, it is probably more appropriate to treat a $2\sigma$ discrepancy in the clustering amplitude as evidence of sample-to-sample fluctuations, than as a signature of new physics beyond the standard paradigm. 

\section{Conclusion}\label{sec6}

The goal of the paper is a comprehensive analysis of the viability of a curvature extension of the standard model of cosmology, achieved by trying to maximise the information from the clustering of galaxies at low redshift. In our approach, this means choosing optimal clustering observables -- the redshift-space distortion parameter $f\sigma_8$ and the Clustering Ratio $\eta_R$ -- and optimising theoretical tools to accurately predict these observables in spatially curved models. Indeed, while the effects of spatial curvature on the kinematics of the Universe are explicitly accounted for by the FLRW metric, the effects of spatial curvature on the clustering and motion of matter is a less explored and far from trivial theoretical issue. 

From a theoretical point of view, we exploit the formalism of Fourier transforms in constant curvature spaces to determine on what scale deviations from Euclidean geometry occur and impact the power spectrum of matter fluctuations. Building on the pioneering work of Matsubara \cite{Matsubara:1999du}, we evaluate how the multipoles of the $2$-point correlation function in redshift space and the Clustering Ratio are affected by spatial curvature and wide-angle effects. First, we revisit Matsubara's formalism to unify it, make it more compact and highlight the physical imprints of curvature. 

We then exploit this formalism to simulate wide-angle and curvature effects on $2$-point statistics, in order to identify the regime of validity of current pipelines for data analysis. In addition, we show that, at least on the scales currently probed by SDSS, it is safe to neglect curvature effects on the Clustering Ratio. As a consequence, this observable retains all its characteristic properties: independence from redshift, galaxy bias and RSD, on linear scales. Simple formulas for implementing curvature corrections are derived which can be applied to analyse future data covering ultra-large volumes.

We also extend the usual (zero-curvature) Alcock-Paczynski corrections to include the effects of curvature in the mapping of the redshift-space distortion and Clustering Ratio observables. Our theoretical analysis produces a new Alcock-Paczynski transformation rule \eqref{qus} for $f \sigma_8$ which significantly deviates from phenomenological formulas adopted in the literature.

By combining the Clustering Ratio and the redshift-space distortion parameter, we extract cosmological information from the perturbations. While the Clustering Ratio is primarily sensitive to the shape of the matter power spectrum, the redshift-space distortion parameter is sensitive to its normalisation and therefore to information about the linear growth rate of structure. This study highlights how these two clustering estimators are affected by parameter degeneracies in quite distinct ways. Their combination is therefore an effective means of nailing down the value of relevant cosmological parameters. 

Traditionally, results from probes of the cosmological background are exploited to help refine constraints on perturbation variables such as $\sigma_{8,0}$ or $f$. In an orthogonal approach, we use probes of the perturbation sector (CR and RSD) to estimate the value of background parameters, in particular $H_0$ and $\omo$. Using a large compilation of data over the redshift range $0 < z \lesssim 2$ and assuming a BBN prior on $\omb$, we show that the clustering constraints, which are completely independent of CMB data, nevertheless rival the CMB in terms of precision, at least for some cosmological parameters, and place interesting limits on the $K\Lambda$CDM extension of the standard $\Lambda$CDM model. Our main findings are:

\begin{itemize}

\item We provide a fully background-independent lower bound on the Hubble constant $H_0>62$ km/s/Mpc at $95\%$ CL ($H_0>73$ km/s/Mpc at $68\%$ CL). This threshold is obtained without assuming flatness or exploiting geometric probes or calibration ladders -- but relying only on the linear measurements of clustering (CR and RSD) and on the primordial abundance of baryons set by BBN measurements (the prior on $\omb h^2$).

\item In the presence of tensions within the standard cosmological paradigm, we find reassuringly that the low-redshift clustering probes alone (without exploiting external information on $\omb$ or $n_s$ or assuming spatial flatness) constrain the matter density to $\omo=0.26\pm 0.04$. This estimate, is in excellent agreement with measurements obtained in a flat $\Lambda$CDM model, from both geometric probes at low redshift and CMB experiments at high redshift.

\item The posterior distributions of parameters for CMB-independent data, i.e. CR + RSD + BBN + BAO + SNIa, indicate that flat $\Lambda$CDM falls within the region containing the true cosmological model with $68\%$ probability. Specifically, we find the tight constraint $\omk=0.0041\,_{-0.0504}^{-0.0500}$.

\item Unlike most low-redshift clustering samples (BAO, full-shape power spectrum, RSD), the CR data set is the only one that does not disagree with the CMB data, given the $K\Lambda$CDM model -- according to Bayesian evidence as measured by DIC statistics. A joint analysis of CR + CMB gives $\omk= -0.023 \pm 0.010$. In terms of betting probability, the spatially curved model is preferred to the flat model with odds of 10\,:\,1, a value too low to reject the flat model without too much risk of being wrong.

\item Using clustering + BBN + BAO information, a CMB-independent measurement of the sound horizon becomes possible even without imposing spatial flatness. We find $r_{\rm d} = 144.57 \pm 2.34~ {\rm Mpc}$ (68\% CL). This is fully compatible with results using CMB anisotropies in a flat $\Lambda$CDM model. It follows that our mean value $H_0=67.55\,_{-1.63}^{+1.60}~ {\rm km}/{\rm s}/{\rm Mpc}$ is in agreement with $H_0$ extrapolated from CMB in flat $\Lambda$CDM \cite{Planck:2018vyg}. This suggests that the CMB-inferred sound horizon may not be responsible for the tensions observed between local and high-redshift $H_0$ determinations.

\end{itemize}

Despite its diversity and heterogeneity, the vast array of data collected in the late Universe is consistent with spatially flat FLRW models. Specifically, our completely CMB-independent analysis of low-redshift data (clustering, BAO, SNIa) does not allow us to reject the flat $\Lambda$CDM scenario embodied by Planck's 6-parameter baseline model. In this perspective, there does not seem to be a new tension related to spatial curvature within the standard model of cosmology. What remains at stake is to understand whether the discordance in the measurements of $\omk$ from the CMB and from low-redshift samples is due to systematic errors or to statistical uncertainties in the Planck data. Future data will help to solve the conundrum. 

We demonstrated that it is very promising to combine the Clustering Ratio with standard probes of clustering such as $f\sigma_8$, in order to resolve degeneracies and increase predictive power in parameter estimations.  Future work will build on this proof of concept
and provide additional measurements of the Clustering Ratio, exploiting the spectroscopic data sets currently available. In particular, we need to construct the covariance matrix to allow the joint analysis of CR + RSD for the same data sets. The scientific gain will be impressive, since we will be able to test non-minimal extensions of the standard model, such as those invoking dynamical dark energy and neutrino phenomenology.

\[\]
\section*{Acknowledgements}

We thank the ICG at the University of Portsmouth for providing computational resources via the Sciama High Performance Compute Cluster. JB and CM are supported by the Institut Physique de l’Univers (IPHU) (Grant No. 013/2020). RM and LP are supported by the South African Radio Astronomy Observatory and the National Research Foundation (Grant No. 75415).

\clearpage
\appendix
\section{AP effect on the RSD parameter $f \sigma_8$}\label{apeffect}

The RSD parameter $f\sigma_8$ is model dependent and determined in a specific {\it fiducial} cosmology. We face the problem of predicting the distortions induced by measuring it in an arbitrary cosmological model other than the true one. This amounts to predicting two effects: geometric distortions from model to model (encapsulated in the AP correction) and clustering distortions (change in the matter power spectrum in different models). The strategy we follow is to study how the wide-angle prediction for the redshift-space multipoles of the 2-point correlation function is biased when estimated in a cosmology other than the true one.

We compute the redshift-space 2-point correlation function in a given cosmology as described in \autoref{sec2.3}. At the same time we need to simulate the redshift space correlation function that would be measured in the fiducial cosmology.

The $2$-point correlations are related to the number density of pairs of objects in a given bin in $\mu=\cos {\gamma_1}$ and $r$. Then the $2$-point correlation function ${\tilde\xi^s_{\rm g}}(\tilde r, \tilde \mu)$ in the fiducial cosmology\footnote{Quantities with a tilde refer to the fiducial model and those without tilde refer to the true model.} is equal to the true $2$-point correlation function ${\xi^s_{\rm g}} ( r, \mu)$ evaluated at a different comoving separation $r=r(\tilde{r}, \tilde{\mu})$ and angle $\mu=\mu(\tilde{r}, \tilde{\mu})$:
\begin{equation}
 {\tilde\xi^s_{\rm g}(\tilde r, \tilde \mu)= \xi^s_{\rm g} ( r, \mu)\,.}
\end{equation}
Given the redshift $z_1$, we compute the fiducial comoving distance $\tilde x_1 = {c}{H_0}^{-1}\int_0^{z_1}{\dif z}/{\tilde E(z)}$. We then use the fact that the angular separation {$\theta$} between position $1$ and $2$ is invariant under a change of cosmological parameters:
\begin{equation}
 \tilde\theta = \theta \,. 
\end{equation}

We use \eqref{chicurved} to express ${\cos\theta}$ as 
\begin{equation}
{\cos\theta} = 
\frac{\tilde C_{12} - \tilde C_1\tilde C_2}{\tilde K \tilde S_1 \tilde S_2} 
\quad\mathrm{ if } \; \tilde K \ne 0 \,,\qquad
{\cos\theta} = 
\frac{\tilde \chi_1^2 + \tilde \chi_2^2 - \tilde\chi^2}{2\tilde\chi_1\tilde\chi_2} 
\quad \mathrm{if} \; \tilde K=0 \,.  
\end{equation}
{From the curved generalisation of the cosine rule:}
\begin{equation}
\tilde S_2^2 = \tilde S_{12}^2\tilde C_1^2 + \tilde S_1^2\tilde C_{12}^2 +\tilde K \tilde S_{12}^2\tilde S_1^2(1-\tilde\mu^2)+2\tilde S_{12}\tilde C_{12}\tilde S_1\tilde C_1\tilde\mu \,,
\end{equation}

from which the radial comoving distance $\tilde x_2 = \tilde a_0 \tilde\chi_2$ and the corresponding redshift $z_2$ follow. Also the positions $x_1$ and $x_2$ can be expressed as a function of $\tilde r$ and $\tilde\mu$. Note that 
\begin{equation}
S_{12}^2 = C_1^2S_2^2 + S_1^2C_2^2 + K S_1^2 S_2^2\,{\sin^2\theta} - 2C_1C_2S_1S_2\,{\cos\theta}.
\end{equation}

It is now possible to compute the functions $\Xi_l^{(n)}(r)$, defined by \eqref{transformln}, in the true cosmology and derive the angular coefficients $c_l^{(n)}$ in \eqref{cnl}. Finally we can use this information to interpolate the values of the comoving distance $r(\tilde r, \tilde \mu)$. In this way we obtain ${\tilde \xi^s_{\rm g}} (\tilde r, \tilde \mu)$ and its multipoles, 
\begin{equation}
\tilde\xi^{(n)} (\tilde r) = \frac{1}{2}\big(2l+1\big)\int_{-1}^{1} \dif \tilde\mu\, L_n(\tilde \mu)\,{\tilde \xi^s_{\rm g}}(\tilde r, \tilde \mu), 
\end{equation}
as in \eqref{2pcfmp}.

We can predict analytically the distortions induced by the AP effect. This is feasible in the plane-parallel or distant-observer approximation, i.e. when it is possible to define radial ($\rpar$, along the line of sight $\hat{\bm{x}}_1$) and transverse ($\rper$) separations. In this regime, the comoving separation is given by $r^2 = {\rpar^2 + \rper^2}$ and $\mu = \rpar/r$. Two quantities are invariant under a change of cosmology: the redshift separation $\Delta z$ along the line-of-sight and the angular separation $\theta$ between $\hat{\bm{x}}_1$ and $\hat{\bm{x}}_2$. This leads to the canonical AP corrections terms
\begin{eqnarray}
\rpar E(z) & = & \tilde r_{\parallel}\tilde E(z)~~\Rightarrow~~\rpar = \apar\tilde{r}_\parallel\,, \\
\frac{\rper}{D_A(z)} & = & \frac{\tilde{r}_\perp}{\tilde D_A(z)}~~\Rightarrow~~ \rper = \aper\tilde{r}_\perp\,,
\end{eqnarray}
where $\apar,\aper$ are the radial and transverse rescaling factors. Then it follows that
\begin{eqnarray}
r = \tilde r\,\aper\left[1 + \big(\lambda^2-1\big)\,\tilde\mu^2\right]^{1/2} \,,\qquad
\mu ={\tilde\mu}\, \lambda\left[ 1 + \big(\lambda^2-1\big)\,\tilde\mu^2 \right]^{-1/2}\quad\text{where}~\lambda=\frac{\apar}{\aper} .
\end{eqnarray}
Expanding in the small parameter $1-\lambda$, we obtain: 
\begin{eqnarray}
r &\simeq& \tilde r \, \aper \bigg [ \bigg ( 1 + \frac{\Delta}{6} - \frac{\Delta^2}{40} \bigg ) L_0(\tilde\mu) + \bigg ( \frac{\Delta}{3} -\frac{\Delta^2}{14} \bigg ) L_2(\tilde\mu) - \frac{\Delta^2}{35}\,L_4(\tilde\mu ) \bigg ] 
\quad\text{where}\quad \Delta=\lambda^2-1\,,~~~
\label{rexp}\\
\mu &\simeq& \tilde \mu\,\big[1 + A(\tilde\mu^2-1)\big]
\quad\text{where}\quad A= (1-\lambda)\Big[ 1+ \frac{3}{8}(1-\lambda) \Big]. \label{muexp}
\end{eqnarray}
Here $\Delta$ and $A$ are small parameters.

In order to understand the main effect of the AP correction we write the $2$-point correlation function explicitly in terms of a truncated Legendre series 
\begin{equation}
\tilde\xi(\tilde r, \tilde \mu) \simeq \xi^{(0)}(r) + \xi^{(2)}(r) L_2(\mu).
\label{xitaprox}
\end{equation}
In the plane-parallel limit the quadrupole is subdominant, so \eqref{xitaprox} should be able to capture most of the radial and angular dependence of the $2$-point correlation function. Since we have shown that it is possible to relate separations and angles in the true cosmology to separations and angles in the fiducial one, we only miss a way to explicitly include this mapping in \eqref{xitaprox}. The most direct way of dealing with this is to expand both the monopole $\xi^{(0)}(r)$ and the quadrupole $\xi^{(2)}(r)$ in Taylor series around an arbitrary scale $r_0$. This is motivated by the fact that at fixed scale $\tilde r$ the value of $r$ is bounded between $r_{\rm min}$ and $r_{\rm max}$ corresponding to $\tilde\mu=0$ and $\tilde\mu=1$ (or the contrary depending on the sign of $\Delta$):

\begin{eqnarray}
\xi^{(l)} (r) & \simeq & \xi^{(l)} (r_0) + \xi^{(l)\prime} (r_0)(r-r_0) +\frac{1}{2} \xi^{(l)\prime\prime} (r_0){ (r-r_0)^2}\quad\text{where}\quad l=0,2\,. \label{mo} 
\end{eqnarray}

In practice, it is enough to truncate \eqref{mo} to first order and to truncate the Legendre expansion \eqref{rexp} at order $2$. {Using \eqref{muexp},} the Legendre polynomial of order $2$ can be expressed as
\begin{equation}
L_2(\mu) = \frac{1}{2}\left \{ 3\tilde\mu^2\left [ (1-A)^2 + 2(1-A)A\tilde\mu^2 + A^2\tilde\mu^4\right ] - 1\right \} ,     
\end{equation}
and neglecting terms in $A^2$ we get the simple Legendre series, 
\begin{equation}
L_2(\mu) \simeq -\frac{2}{5}A\,L_0({\tilde\mu}) + \Big(1-\frac{2}{7}A\Big)L_2({\tilde\mu}) + \frac{24}{35}A\,L_4({\tilde\mu}).
\end{equation}

A natural choice for $r_0$ in \eqref{mo} is $r_0 = \tilde r\aper (1 + \Delta/6)$, so that $r-r_0 \simeq \tilde r\,\aper \,{\Delta}\, L_2(\mu)/3$. In this way the most important term contributing to the monopole will be $\xi^{(0)}(r_0)$ where $r_0 = \alpha\, \tilde r$ and $\alpha\simeq \aper (1+\Delta/6)$. Note that $\alpha$ in this approaximation agrees with the volume factor defined in \eqref{eq:def_alpha}. 
However this is not the only term contributing to the monopole in the product $\xi^{(2)}(r)L_2(\mu)$. There are two additional contributions: 
\begin{eqnarray}
\tilde\xi^{(0)}(\tilde r)&\simeq& \xi^{(0)}(\alpha\, \tilde r) -\frac{2}{5}A\,\xi^{(2)}(r_0) + {\tilde r \aper}\,\frac{\Delta}{15} \Big(1-\frac{2}{7}\,A \Big)\,\xi^{(2)\prime}(r_0)
\label{monoap}\\
{\alpha}&\simeq& {\aper \Big(1+\frac{\Delta}{6}\Big)\simeq 
\big( \alpha_\parallel\alpha_\perp^2 \big)^{1/3}}.
\end{eqnarray}

\begin{figure*}
\centering
\includegraphics[clip, trim = 2.3cm 7.1cm 2.2cm 7.6cm, width=0.46\linewidth]{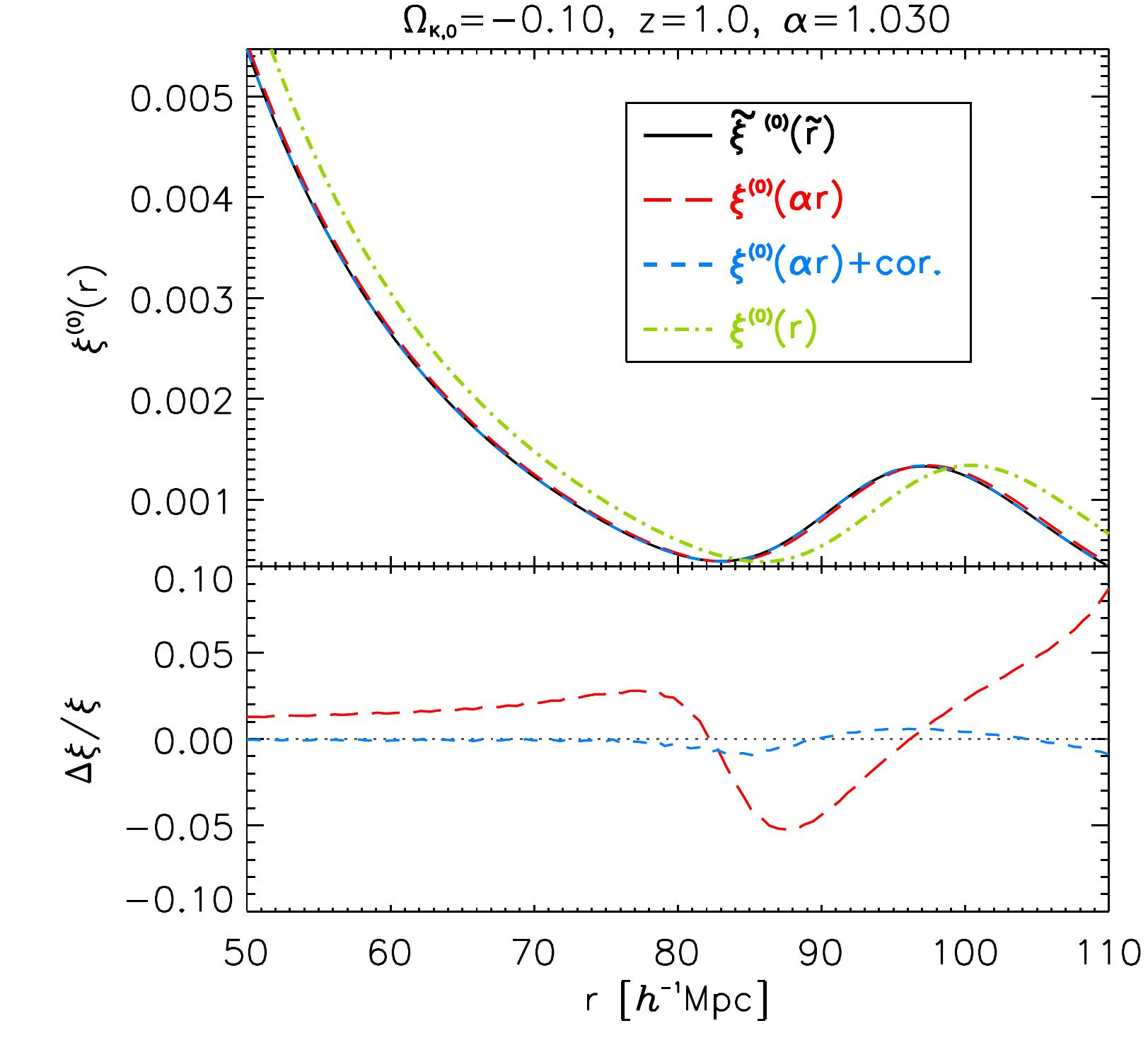}
\hskip4mm
\includegraphics[clip, trim = 2.3cm 7.1cm 2.2cm 7.6cm, width=0.46\linewidth]{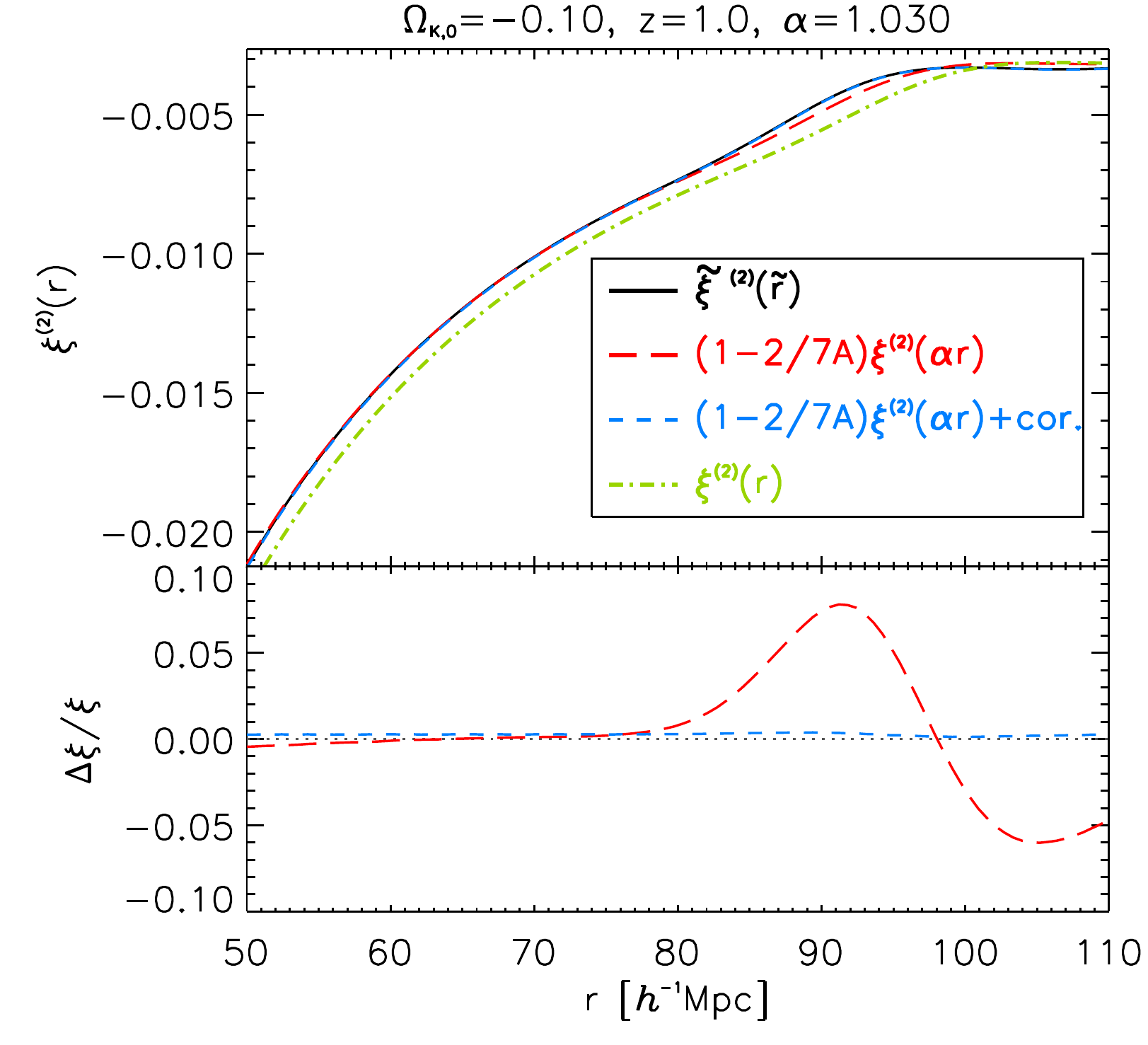}
\caption{\small {\it Top:} AP effect on the monopole \eqref{monoap} (left) and quadrupole \eqref{a16} (right). Solid black line shows the true distorted multipoles. Red long-dashed line shows the leading (first) contribution and blue short-dashed line is the correction. Green dot-dashed line shows the multipole without AP effect. Fiducial model: $\omo = 0.37$, $\omk=0$; true model: $\omk=-0.1$, $\omo=0.32$. {\it Bottom:} Fractional difference relative to true distorted multipoles.} 
\label{fig:ap}
\end{figure*}

This shows that the intrinsic anisotropy, i.e. the quadrupole, will leak into the AP correction to the monopole. An explicit comparison is shown in the left panel of \autoref{fig:ap}, where we choose a true cosmology with $\omk=-0.1$ and $\omo=0.32$ (if distances are in $h^{-1}$Mpc, then $H_0$ does not matter). The fiducial cosmology is a flat $\Lambda$CDM with $\omo=0.37$. In the same plot, we compare the $2$-point correlation function (which includes wide-angle and curvature effects), numerically computed in the fiducial cosmology, with that computed in the true cosmology. It is clear that the main effect is a shift in scale ($\alpha\, {\tilde r}$), but it remains a mismatch of $\sim 1-5$\%, which is explained by the correction in \eqref{monoap}.

Using the same reasoning, one obtains for the quadrupole
\begin{eqnarray}
\tilde\xi^{(2)}(\tilde r) &\simeq & \xi^{(2)}(\alpha\, \tilde r)\Big( 1-\frac{2}{7}\,A \Big) + {1\over3} \xi^{(0)\prime}(r_0)\tilde r \aper \Delta 
+ \xi^{(2)\prime}(r_0)\Big[ \frac{2}{147}\tilde r\aper\Delta (7 - 2 A) + {16\over 245}\tilde r \Delta A \aper\! \Big]\! ,~~~~~~ \label{a16}
\end{eqnarray}

where the dominant term is the first one. The leading and correction terms are compared to the distorted quadrupole in the right panel of \autoref{fig:ap}. The fact that the correction terms are sufficient to describe the distorted $2$-point correlation function at the sub-percent level validates the hypothesis of a sub-dominant quadrupole via the ansatz \eqref{xitaprox}. Note that similar expressions have been obtained by \cite{Xu:2012fw} and we checked that they match ours at linear order in $\Delta$. 

Equation \eqref{monoap} shows that the dominant term in the monopole does not affect its amplitude, but rather results in a change of scale given by $\alpha\, \tilde r$. As a result, fixing the value of the monopole with data allows us to constrain the shift scale which needs to be used in the quadrupole. By contrast, the dominant term in the quadrupole \eqref{a16} is not only shifted but also undergoes a change in amplitude by a factor $(1-2A/7)$. As a result we obtain the transformation formula 
\begin{equation}
\widetilde {f\sigma}_8 \simeq \left(1-\frac{2}{7}A\right) f\sigma_8 , 
\end{equation}
which allows us to map any given value of the RSD factor $f\sigma_8$ to the value one would have measured in a fiducial cosmology.

\section{AP effect on the Clustering Ratio ${\eta_R}$}\label{apcr}

As with the RSD parameter, the Clustering Ratio is a model-dependent quantity that is estimated from the data in a given fiducial cosmology and then used in a likelihood analysis to identify a best fitting cosmological model. We are thus faced with the similar problem of relating the theoretical value it has in an arbitrary model to the value that would be measured in the fiducial model.

\begin{figure}[!]
\centering
\includegraphics[clip, trim = 0cm 17cm 0cm 0.9cm, width=0.9\linewidth]{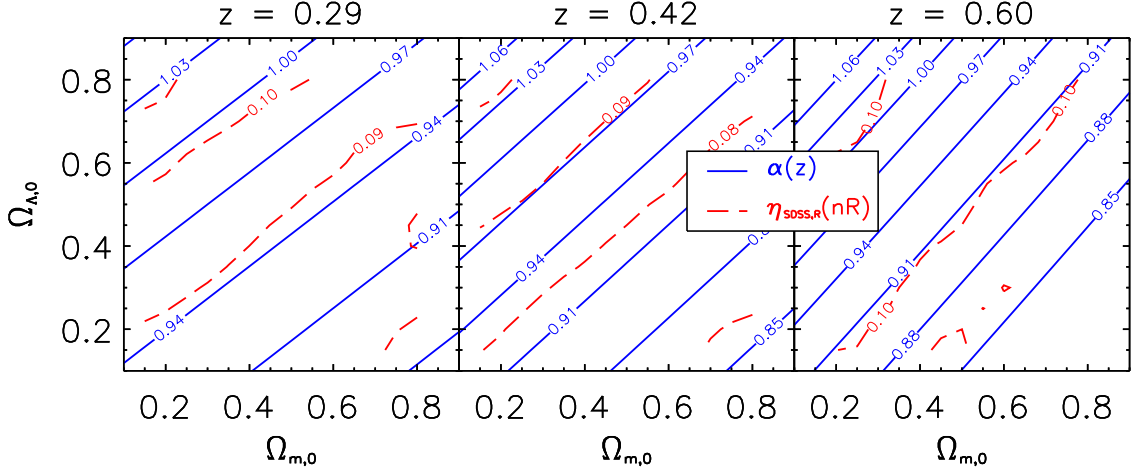}
\caption{Red long-dashed lines show 3 measurements (left to right panels) of $\eta_R(nR)$
 ($n=2.1,R=22h^{-1}$Mpc) in SDSS as a function of the fiducial cosmology
 ($\Omega_{m,0}$, $\Omega_{\Lambda,0}$). Solid blue lines give 
the rescaling parameter $\alpha$ used in the AP correction.} 
\label{fig:etaap}
\end{figure}

\autoref{fig:etaap} illustrates the cosmological dependence of the Clustering Ratio of galaxies estimated in the SDSS (DR7 and DR12) redshift survey. Following \autoref{apeffect}, we neglect wide-angle effects and make a plane-parallel approximation. The idea is to express the smoothed matter correlation function $\xi_R(r)$ as a function of $\xi(r)$, using \eqref{xismooth}. We start with the identities
\begin{eqnarray}
 \xi^{(n)}(r) & = & 4\pi \int_0^{\infty}\dif k\,k^2\, P^{(n)}(k)\, j_n(kr) , \\
 P^{(n)}(k) & = & \frac{1}{2\pi^2}\int_0^{\infty}\dif r\, r^2\,\xi^{(n)}(r)\,j_n(kr) .
\end{eqnarray}

Then we can express the multipoles of the smoothed correlation function as
\begin{eqnarray}
 \xi_R^{(n)}(r) &=& \frac{1}{\pi r} \int_0^{\infty}\dif r'\, r'\, \xi^{(n)}(r')\, J_n(r, r', R) ,
 \label{xirbn}\\
 J_n(r, r', R) &\equiv& \frac{3\pi}{2R^3} \int_{|r-r'|}^{2R}\dif x\, x\, L_n\bigg( \frac{r^2+r'^2-x^2}{2r r'} \bigg) \bigg[ 1 - \frac{3}{4} \frac{x}{R} + \frac{1}{2}\bigg( \frac{x}{2R} \bigg)^2\, \bigg] .
\label{jkernel}
\end{eqnarray}

It follows that, when the monopole is considered, $J_n$ depends only on the radius $R$ and on $|r-r'|$. Therefore we can express $r\xi_R^{(0)}(r)$ as a convolution of $r\xi^{(0)}(r)$ with an appropriate kernel $w$:
\begin{eqnarray}
 r\xi_R^{(0)}(r) &=& \int_{r-2R}^{r+2R} \dif r'\, r'\,\xi^{(0)}(r')\, w\bigg( \frac{|r-r'|}{R} \bigg),
 \label{rximono}
 \\
 w(x) &\equiv& \frac{3}{5R} \bigg[ 1 - \frac{5}{4} x^2 + \frac{5}{8}x^3 - \bigg ( \frac{x}{2} \bigg)^5\, \bigg].
\label{wkernel}
\end{eqnarray}

By \eqref{monoap}, $\tilde\xi^{(0)}( \tilde r) \simeq \xi^{(0)}(\alpha \tilde r)$ and \eqref{rximono} shows that the same relation applies to the smoothed correlation function:
\begin{equation}
\tilde \xi^{(0)}_{\tilde R}(\tilde r) = \xi_{\alpha \tilde R}^{(0)}(\alpha \tilde r) .
\end{equation}
Following the same reasoning, the variance $\sigma_R^2$ is obtained as 
\begin{equation}
 \sigma_R^2 = \frac{3}{R^3} \int_0^{2R} \dif r'\, r^{\prime 2}\,\xi^{(0)}(r') \bigg[ 1 - \frac{3}{4} \frac{r'}{R} + \frac{1}{2}\bigg( \frac{r'}{2R} \bigg)^3\, \bigg] .
 \label{sigmar}
\end{equation}
Therefore the same mapping applies:
\begin{equation}
\tilde \sigma_{\tilde R}^2 = \sigma_{R}^2 . 
\end{equation}
Since the mapping from true cosmology to fiducial is the same for both the smoothed $2$-point correlation and the variance, we deduce that the Clustering Ratio transforms as
\begin{equation}
\tilde \eta_{\tilde R}(\tilde r) = \eta_R(\alpha \tilde r). 
\end{equation} 
As a matter of fact, the Clustering Ratio should be invariant (with respect to the background cosmology) when $\alpha$ is constant. Inspecting \autoref{fig:etaap}, we see that at each redshift of the SDSS survey, the cosmological dependence of the Clustering Ratio is the same as the cosmological dependence of $\alpha$. This confirms that we can apply the AP correction for the Clustering Ratio in the same way as for the $2$-point correlation function.

\clearpage
\bibliography{References}
\bibliographystyle{JHEP}

\end{document}